\documentclass[aps,pra,preprint,superscriptaddress,longbibliography]{revtex4-2}

\usepackage{graphicx}
\usepackage{amsmath}
\usepackage{txfonts}
\usepackage{xcolor}
\usepackage{amsfonts}
\usepackage{amssymb}
\usepackage{comment}
\usepackage{cancel}
\usepackage[pdftex]{hyperref}

\begin{document}

	\title{Affine spin-pencil factorization of generalized driven anisotropic Rabi--Stark Hamiltonians in an inhomogeneous orthosymplectic superalgebra}

	\author{B.~M. Rodr\'iguez-Lara} 
	\email[e-mail: ]{blas.rodriguez@gmail.com}
	\affiliation{Universidad Polit{\'e}cnica Metropolitana de Hidalgo, Tolcayuca, Hidalgo, M{\'e}xico}
	
	\author{F.~H. Maldonado-Villamizar}
	\email[e-mail: ]{fmaldonado@inaoep.mx}
	\affiliation{SECIHTI-Instituto Nacional de Astrof\'{i}sica, \'{O}ptica y Electr\'{o}nica, Calle Luis Enrique Erro No. 1. Sta. Ma. Tonantzintla, Pue. C.P. 72840, Mexico}
	
	\author{A. Kafuri} 
	\email[email: ]{anuar.kafuri@cinvestav.mx}
	\affiliation{Centro de Investigación y Estudios Avanzados del Instituto Politécnico Nacional}
	
	\author{A. Moroz} 
	\email[email: ]{wavescattering@yahoo.com}
	\affiliation{Wave-scattering.com}

	\begin{abstract}
		We construct a supersymmetric factorization for generalized driven anisotropic Rabi--Stark Hamiltonians using an affine factor $\hat{A}=\hat{C}+\hat{a}\hat{D}$ in an inhomogeneous orthosymplectic superalgebra, with two real invertible spin matrices $\hat{C}$ and $\hat{D}$.
		The six real pencil parameters constrain nine Hamiltonian frequencies and one energy offset up to an overall frequency scale.
		The normal product $\hat{A}^{\dagger}\hat{A}$ has an exact zero-energy ground doublet and, with its antinormal partner $\hat{A}\hat{A}^{\dagger}$, gives a Witten index of two.
		For distinct pencil roots with finite baseline slopes, coherent gauges define a Bargmann hierarchy and endpoint conditions for exceptional finite-polynomial states.
		The spin grading organizes four limiting cases, comprising two driven-oscillator families with every baseline exact and the constrained anisotropic-Rabi and Rabi--Stark sheets, which meet at the degenerate-qubit isotropic Rabi model.
		An aligned interior example adds direct and longitudinal drives, Stark deformation, and intensity-dependent coupling while retaining the factorization.
	\end{abstract}
	
	\maketitle
	\newpage
	
	\section{Introduction}
	
	Rabi's driven two-level system is the starting point for coherent light--matter dynamics.
	A magnetic moment in a static field driven by a transverse oscillatory field~\cite{Rabi1936p324,Rabi1937p652},
	\begin{align}
		\frac{\hat{H}_{\mathrm{R}}(t)}{\hbar} =&~ \frac{1}{2}\omega_{0}\hat{\sigma}_{z} + \Omega\cos \omega t \, \hat{\sigma}_{x},
	\end{align}
	has transition probabilities set by the drive frequency $\omega$, amplitude $\Omega$, and detuning $\omega_{0} - \omega$.
	This resonance mechanism became the working principle of magnetic resonance spectroscopy~\cite{Bloch1946p127,Purcell1946p37}, the phase reference behind Ramsey spectroscopy and atomic clocks~\cite{Ramsey1950p695}, and the two-level transition language of masers and lasers~\cite{Gordon1955p1264,Schawlow1958p1940,Maiman1960p493}.
	
	Jaynes and Cummings quantized the field as a single radiation mode and, under the rotating-wave approximation, obtained the first fully quantum solvable single-mode two-level model~\cite{Jaynes1963p89},
	\begin{align}
		\frac{\hat{H}_{\mathrm{JC}}}{\hbar} =&~ \omega\hat{a}^{\dagger}\hat{a} + \frac{1}{2}\omega_{0}\hat{\sigma}_{z} + g \left( \hat{a}\hat{\sigma}_{+} + \hat{a}^{\dagger}\hat{\sigma}_{-} \right),
	\end{align}
	where total excitation-number conservation, $[\hat{H}_{\mathrm{JC}},\hat{N}] = 0$ with $\hat{N} = \hat{a}^{\dagger} \hat{a} + (1 + \hat{\sigma}_{z})/2$, partitions the infinite Hilbert space into a one-dimensional ground-state block and independent two-dimensional excitation blocks.
	Exact solvability means that the closed-form diagonalization of the finite blocks provides the complete spectrum and eigenstates.
	
	The full quantum Rabi model,
	\begin{align}
		\frac{\hat{H}_{\mathrm{QR}}}{\hbar} =&~ \omega\hat{a}^{\dagger}\hat{a} + \frac{1}{2}\omega_{0}\hat{\sigma}_{z} + g\left( \hat{a}^{\dagger} + \hat{a} \right)\hat{\sigma}_{x},
	\end{align}
	does not conserve the total excitation number, $[\hat{H}_{\mathrm{QR}},\hat{N}] \neq 0$, but preserves parity, $[\hat{H}_{\mathrm{QR}},\hat{\Pi}] = 0$ with $\hat{\Pi}=e^{i\pi\hat{a}^{\dagger}\hat{a}}\hat{\sigma}_{z}$.
	The counter-rotating terms, $( \hat{a}\hat{\sigma}_{-} + \hat{a}^{\dagger} \hat{\sigma}_{+} )$, change the total excitation number by two units, linking the Jaynes--Cummings excitation blocks into two infinite parity ladders.
	Quasi-exact solvability in the quantum Rabi model, which determines selected eigenvalues and eigenstates analytically while leaving the complete spectrum undetermined, appeared first when Judd found finite polynomial eigenstates at exceptional parameter values~\cite{Judd1979p1685}, and Ku{\'s} later organized these points through constraint-polynomial conditions~\cite{Kus1985p2792}.
	Full exact solvability gives a complete analytic characterization of the spectrum and eigenstates without requiring closed-form eigenvalue formulas.
	Braak gave such a global parity-resolved construction in Bargmann space, where zeros of spectral $G$-functions determine the regular spectrum~\cite{Braak2011p100401}.
	
	The Jaynes--Cummings model became the algebraic template for solvable bosonic-mode--pseudospin Hamiltonians.
	It admits a dynamical-superalgebra organization~\cite{Buzano1989p137}, supersymmetric partner sectors~\cite{Andreev1989p507,Lee1994pR4}, and ladder operators with coherent states adapted to the rotating-wave spectrum~\cite{Hussin2005p122102}.
	Supersymmetric extensions based on graded Lie algebras produced analytic spectra for Jaynes--Cummings-type Hamiltonians~\cite{Alhaidari2006p15391}, while generalized models with nonlinear boson terms, intensity-dependent couplings, and multiboson exchange preserved the same operator-diagonalization mechanism~\cite{MaldonadoVillamizar2021p16467}.
	Recent supersymmetric treatments showed the interchangeability between Jaynes--Cummings and anti-Jaynes--Cummings dynamics and used this connection to compute observables and full photon-counting statistics~\cite{BocanegraGaray2024p043218}.
	
	This algebraic viewpoint extends to quantum Rabi models, where the $\mathrm{osp}(2|2)$ superalgebra provides the underlying graded structure~\cite{Schmitt1990p305}.
	Factorization gives the simplest constructive use of that structure.
	Writing a Hamiltonian as the normal- or antinormal-ordered product of a first-order bosonic--pseudospin operator and its adjoint gives an exact analytic ground state, fixes its energy algebraically, and identifies baseline energies for the rest of the spectrum.
	Intensity-dependent Rabi Hamiltonians realize this mechanism through $\mathrm{su}(1,1)$ reductions and supersymmetric partner factorizations~\cite{RodriguezLara2014p1719}, while anisotropic Rabi Hamiltonians realize it on supersymmetric parameter curves with exact factorizable states~\cite{Tomka2014p063839,Tomka2015p13097}.
	Our recent work used linear combinations of $\mathrm{osp}(2|2)$ generators to produce a family of anisotropic Rabi--Stark Hamiltonians with an exact unique ground state~\cite{Kafuri2024pC82}.
	The anisotropic quantum Rabi model is physically relevant in platforms where rotating and counter-rotating channels can be controlled independently, including trapped-ion simulations based on detuned red and blue sidebands~\cite{RodriguezLara2005p023811,Pedernales2015p15472} and circuit-QED frequency-modulation schemes~\cite{Wang2018p053061,Wang2019p4569}.
	
	Here we show that these factorization mechanisms are corners of a single affine regular spin-pencil construction.
	We embed the anisotropic Rabi algebra in an inhomogeneous orthosymplectic superalgebra in Sec.~\ref{sec:Sec2}. 
	In Sec.~\ref{sec:Sec3}, we introduce an affine spin-sector pencil evaluated at the annihilation operator and use its normal product to generate a constrained family of generalized driven anisotropic Rabi--Stark Hamiltonians. 
	We show that the standard matrix of the pencil controls the coherent-spinor zero modes, while a Bargmann coherent-gauge construction gives baseline energies and endpoint compatibility conditions for finite-polynomial states in Sec.~\ref{sec:Sec4}.
	In Sec.~\ref{sec:Sec5}, we use the spin-inversion grading to organize the identity--identity, exchange--identity, identity--exchange, and exchange--exchange corners, and an aligned interior example shows how direct drives, Stark terms, longitudinal spin-dependent drives, and intensity-dependent transverse spin drives enter without leaving our factorization framework.
	We close with a summary and our conclusions in Sec.~\ref{sec:Sec6}.
	
	\section{Inhomogeneous orthosymplectic superalgebra}
	\label{sec:Sec2}
	
	We start from the anisotropic quantum Rabi model~\cite{Xie2014p021046},
	\begin{align}
		\frac{\hat{H}_{\mathrm{AR}}}{\hbar} =&~ \omega \hat{a}^{\dagger} \hat{a} +\frac{1}{2}\omega_{0}\hat{\sigma}_{z} + g_{1} \left( \hat{a}^{\dagger} \hat{\sigma}_{-} + \hat{a} \hat{\sigma}_{+} \right) + g_{2}\left( \hat{a}^{\dagger} \hat{\sigma}_{+} + \hat{a} \hat{\sigma}_{-} \right),
	\end{align}
	describing a bosonic mode of frequency $\omega$, with creation and annihilation operators $\hat{a}^{\dagger}$ and $\hat{a}$, coupled to a two-level system of transition frequency $\omega_{0}$, with raising and lowering operators $\hat{\sigma}_{\pm}$, through Jaynes--Cummings and anti--Jaynes--Cummings channels with coupling strengths $g_{1}$ and $g_{2}$.
	We refer to this two-level system as the spin.
	The model preserves the parity underlying the exact-solvability structure of the quantum Rabi model~\cite{Braak2011p100401,Braak2019p1259},
	\begin{align}
		\hat{\Pi} =&~ e^{i \pi \hat{a}^{\dagger} \hat{a}} \hat{\sigma}_{z}, \qquad
		\left[ \hat{H}_{\mathrm{AR}}, \hat{\Pi} \right] = 0.
	\end{align}
	
	The bosonic mode ladder operators with the identity realize the Heisenberg--Weyl Lie algebra $\mathrm{hw}(1)$,
	\begin{align}
		\left[ \hat{a}, \hat{a}^{\dagger} \right] =&~ \hat{I},
	\end{align}
	while the spin Cartan and ladder operators realize the $\mathrm{su}(2)$ Lie algebra, 
	\begin{align}
		\left[ \hat{\sigma}_{z}, \hat{\sigma}_{\pm} \right] =&~ \pm 2 \hat{\sigma}_{\pm}, \qquad
		\left[ \hat{\sigma}_{+}, \hat{\sigma}_{-} \right] =~\hat{\sigma}_{z}.
	\end{align}
	The combined set of bosonic and spin generators realizes the direct-sum Lie algebra $\mathrm{hw}(1) \oplus \mathrm{su}(2)$ under the commutator, $[\hat{X}, \hat{Y}] = \hat{X} \hat{Y} - \hat{Y} \hat{X}$.
	The spin ladder operators additionally close with the identity under the anticommutator,
	\begin{align}
		\left\{ \hat{\sigma}_{+}, \hat{\sigma}_{-} \right\} =&~ \hat{I},
	\end{align}
	with $\left\{ \hat{X}, \hat{Y} \right\} = \hat{X} \hat{Y} + \hat{Y} \hat{X}$.
	The simultaneous commutator and anticommutator closures organize the generators into a $\mathbb{Z}_{2}$-graded algebra through the supercommutator
	\begin{align}
		\left[ \hat{X}, \hat{Y} \right]_{\mathrm{s}} =&~ \hat{X} \hat{Y} - (-1)^{\lvert \hat{X} \rvert \lvert \hat{Y} \rvert} \hat{Y} \hat{X}, \qquad
		\lvert \hat{X} \rvert, \lvert \hat{Y} \rvert \in \left\{ 0,1 \right\},
	\end{align}
	where we assign the bosonic ladder operators and the identity an even grade,
	\begin{align}
		\lvert \hat{I} \rvert =&~ \lvert \hat{a} \rvert = \lvert \hat{a}^{\dagger} \rvert = 0,
	\end{align}
	and an odd grade to the spin ladder operators,
	\begin{align}
		\lvert \hat{\sigma}_{\pm} \rvert =&~ 1.
	\end{align}
	The first-order set $\{ \hat{I}, \hat{a}^{\dagger}, \hat{a}, \hat{\sigma}_{\pm} \}$ realizes the Heisenberg--Weyl Lie superalgebra $\mathrm{hw}(1|1)$~\cite{Jafarov2013p103506},
	\begin{align}
		\left[ \hat{a}, \hat{a}^{\dagger} \right]_{\mathrm{s}} = \hat{I}, \qquad
		\left[ \hat{\sigma}_{+}, \hat{\sigma}_{-} \right]_{\mathrm{s}} = \hat{I},
	\end{align}
	with all other supercommutators vanishing.
	
	Quadratic products of the first-order generators inherit the $\mathbb{Z}_{2}$ grading,
	\begin{align}
		\lvert \hat{X} \hat{Y} \rvert =&~ \left( \lvert \hat{X} \rvert + \lvert \hat{Y} \rvert \right) \bmod 2.
	\end{align}
	Pure bosonic and pure spin products are even graded, while mixed bosonic--spin products are odd graded.
	The independent pure bosonic quadratic products,
	\begin{align}
		\hat{K}_{0} =&~ \frac{1}{4} \left\{ \hat{a}^{\dagger}, \hat{a} \right\}, \qquad
		\hat{K}_{+} = \frac{1}{2} \hat{a}^{\dagger 2}, \qquad
		\hat{K}_{-} = \frac{1}{2} \hat{a}^{2},
	\end{align}
	realize the $\mathrm{sp}(2,\mathbb{R}) \simeq \mathrm{su}(1,1)$ Lie algebra, 
	\begin{align}
		\left[\hat{K}_{0}, \hat{K}_{\pm} \right]_{\mathrm{s}} =&~ \pm \hat{K}_{\pm}, \qquad
		\left[\hat{K}_{+}, \hat{K}_{-} \right]_{\mathrm{s}} = - 2 \hat{K}_{0}, 
	\end{align}
	while the pure spin products reduce to the identity and the spin-inversion operator,
	\begin{align}
		\hat{\sigma}_{+} \hat{\sigma}_{-} =&~ \frac{1}{2}\hat{I} + \hat{B}, \qquad
		\hat{\sigma}_{-} \hat{\sigma}_{+} = \frac{1}{2}\hat{I} - \hat{B}, \qquad
		\hat{B} = \frac{1}{2}\hat{\sigma}_{z},
	\end{align}
	as the spin ladder operators are nilpotent $\hat{\sigma}_{\pm}^{2} = 0$.
	The spin-inversion operator $\hat{B}$ generates the $\mathrm{so}(2) \simeq \mathrm{u}(1)$ Lie algebra.
	These noncentral pure quadratic generators realize the even-grade sector of an orthosymplectic superalgebra,
	\begin{align}
		\mathrm{osp}(2|2)_{\bar{0}} =&~ \mathrm{sp}(2,\mathbb{R}) \oplus \mathrm{so}(2) \simeq \mathrm{su}(1,1) \oplus \mathrm{u}(1).
	\end{align}
	
	The mixed bosonic--spin quadratic products are odd graded and separate into the Jaynes--Cummings~\cite{MaldonadoVillamizar2021p16467} and anti--Jaynes--Cummings~\cite{Kafuri2024pC82} generators,
	\begin{align}
		\begin{aligned}
			\hat{Q}_{+} =&~ \hat{a} \hat{\sigma}_{+}, \qquad
			&\hat{Q}_{-} =&~ \hat{a}^{\dagger} \hat{\sigma}_{-}, \\
			\hat{R}_{+} =&~ \hat{a} \hat{\sigma}_{-}, \qquad
			&\hat{R}_{-} =&~ \hat{a}^{\dagger} \hat{\sigma}_{+},
		\end{aligned}
	\end{align}
	with supercommutators
	\begin{align}
		\begin{aligned}
			\left[ \hat{Q}_{+}, \hat{Q}_{-} \right]_{\mathrm{s}} =&~ 2 \hat{K}_{0} + \hat{B} \equiv \hat{H}_{Q}, \\
			\left[ \hat{R}_{+}, \hat{R}_{-} \right]_{\mathrm{s}} =&~ 2 \hat{K}_{0} - \hat{B} \equiv \hat{H}_{R},
		\end{aligned}
	\end{align}
	providing the even-grade central elements of the Jaynes--Cummings and anti--Jaynes--Cummings subalgebras,
	\begin{align}
		\begin{aligned}
			\left[ \hat{H}_{Q}, \hat{Q}_{\pm} \right]_{\mathrm{s}} =&~ 0,\qquad 
			\left[ \hat{H}_{R}, \hat{R}_{\pm} \right]_{\mathrm{s}} = 0,
		\end{aligned}
	\end{align}
	while the spin-inversion operator closes the even--odd supercommutators,
	\begin{align}
		\begin{aligned}
			\left[ \hat{B}, \hat{Q}_{\pm} \right]_{\mathrm{s}} =&~ \pm \hat{Q}_{\pm},\qquad
			\left[ \hat{B}, \hat{R}_{\pm} \right]_{\mathrm{s}} = \mp \hat{R}_{\pm},
		\end{aligned}
	\end{align}
	realizing the Jaynes--Cummings $\mathrm{u}_{Q}(1|1)$ and the anti--Jaynes--Cummings $\mathrm{u}_{R}(1|1)$ Lie superalgebras.
	The Jaynes--Cummings and anti--Jaynes--Cummings generators $\{ \hat{Q}_{\pm}, \hat{R}_{\pm} \}$ realize the odd-grade sector $\mathrm{osp}(2|2)_{\bar{1}}$ of the orthosymplectic superalgebra~\cite{Schmitt1990p305,MaldonadoVillamizar2021p16467,Kafuri2024pC82},
	\begin{align}
		\mathrm{osp}(2|2) =&~ \mathrm{osp}(2|2)_{\bar{0}} \oplus \mathrm{osp}(2|2)_{\bar{1}}.
	\end{align}
	The Jaynes--Cummings and anti--Jaynes--Cummings subalgebras couple through the cross supercommutators
	\begin{align}
		\left[ \hat{Q}_{+}, \hat{R}_{+} \right]_{\mathrm{s}} =&~ 2 \hat{K}_{-}, \qquad
		\left[ \hat{Q}_{-}, \hat{R}_{-} \right]_{\mathrm{s}} = 2 \hat{K}_{+},
	\end{align}
	while the even generators act on the odd sector through
	\begin{align}
		\begin{aligned}
			\left[ \hat{K}_{0}, \hat{Q}_{\pm} \right]_{\mathrm{s}} =&~ \mp \frac{1}{2} \hat{Q}_{\pm}, \qquad
			&\left[ \hat{K}_{0}, \hat{R}_{\pm} \right]_{\mathrm{s}} =&~ \mp \frac{1}{2} \hat{R}_{\pm}, \\
			\left[ \hat{K}_{+}, \hat{Q}_{+} \right]_{\mathrm{s}} =&~ -\hat{R}_{-}, \qquad
			&\left[ \hat{K}_{+}, \hat{R}_{+} \right]_{\mathrm{s}} =&~ -\hat{Q}_{-}, \\
			\left[ \hat{K}_{-}, \hat{Q}_{-} \right]_{\mathrm{s}} =&~ \hat{R}_{+}, \qquad
			&\left[ \hat{K}_{-}, \hat{R}_{-} \right]_{\mathrm{s}} =&~ \hat{Q}_{+},
		\end{aligned}
	\end{align}
	with the remaining supercommutators vanishing, completing the graded closure
	\begin{align}
		\left[ \hat{X}_{\bar{i}}, \hat{Y}_{\bar{j}} \right]_{\mathrm{s}} \in&~ \mathrm{osp}(2|2)_{\overline{i+j}}, \qquad
		\hat{X}_{\bar{i}} \in \mathrm{osp}(2|2)_{\bar{i}}, \qquad
		\hat{Y}_{\bar{j}} \in \mathrm{osp}(2|2)_{\bar{j}}
	\end{align}
	with $i,j \in \{0,1\}$ and addition modulo two.
	
	The first-order Heisenberg--Weyl generators extend the homogeneous orthosymplectic superalgebra by translations, realizing the inhomogeneous orthosymplectic Lie superalgebra~\cite{AlvarezMoraga2004p179},
	\begin{align}
		\mathrm{hosp}(2|2) =&~ \mathrm{osp}(2|2) \ltimes \mathrm{hw}(1|1).
	\end{align}
	The homogeneous orthosymplectic generators preserve the linear span of the first-order generators,
	\begin{align}
		\left[ \hat{X}, \hat{Y} \right]_{\mathrm{s}} \in&~ \mathrm{hw}(1|1), \qquad 
		\hat{X} \in \mathrm{osp}(2|2), \qquad   
		\hat{Y} \in \mathrm{hw}(1|1),
	\end{align}
	so the Heisenberg--Weyl superalgebra forms an ideal of the inhomogeneous algebra,
	\begin{align}
		\mathrm{hw}(1|1) \triangleleft&~ \mathrm{hosp}(2|2).
	\end{align}
	Taking the quotient by the translation ideal recovers the homogeneous orthosymplectic superalgebra,
	\begin{align}
		\mathrm{hosp}(2|2)/\mathrm{hw}(1|1) \simeq&~ \mathrm{osp}(2|2).
	\end{align}
	This decomposition has a Levi-like form with a solvable Heisenberg--Weyl radical and a homogeneous orthosymplectic factor~\cite{Humphreys1972}.
	The anisotropic quantum Rabi Hamiltonian is a linear element of the homogeneous orthosymplectic algebra up to an additive scalar,
	\begin{align}
		\frac{\hat{H}_{\mathrm{AR}}}{\hbar} =&~ 2 \omega \hat{K}_{0} + \omega_{0} \hat{B} + g_{1} \left( \hat{Q}_{+} + \hat{Q}_{-} \right) + g_{2} \left( \hat{R}_{+} + \hat{R}_{-} \right) - \frac{1}{2} \omega.
	\end{align} 
	Parity-breaking driven extensions with first-order bosonic translations or spin flips lie in the inhomogeneous $\mathrm{hosp}(2|2)$ layer.
	Normal and antinormal products of these generators are quadratic combinations and belong to the corresponding universal enveloping algebra.
	
	\section{Regular spin-pencil factorization}
	\label{sec:Sec3}
	The inhomogeneous orthosymplectic superalgebra motivates us to introduce a dimensionless spin pencil~\cite{Gantmacher1998},
	\begin{align}
		\hat{P}(z) =&~ \hat{C} + z\hat{D}, \qquad
		\det \hat{P}(z) \not\equiv 0,
	\end{align}
	whose evaluation at the annihilation operator gives the affine first-order factor,
	\begin{align}
		\hat{A} =&~ \hat{P}(\hat{a}) = \hat{C} + \hat{a}\hat{D} = c_{0} + c_{-}\hat{\sigma}_{-} + c_{+}\hat{\sigma}_{+} + d_{0}\hat{a} + d_{-}\hat{R}_{+} + d_{+}\hat{Q}_{+}.
	\end{align}
	We take real, dimensionless, and invertible spin matrices,
	\begin{align}
		\hat{\Xi} =&~ \xi_{0} + \xi_{-}\hat{\sigma}_{-} + \xi_{+}\hat{\sigma}_{+}, \qquad
		\xi_{0}^{2} - \xi_{+}\xi_{-} \neq 0, \qquad
		\xi_{j} \in \mathbb{R}, \qquad
		(\hat{\Xi},\xi) \in \left\{ (\hat{C},c),(\hat{D},d) \right\}.
		\label{CDmatrices}
	\end{align}
	The invertibility of $\hat{C}$ excludes zero pencil eigenvalues, while the invertibility of $\hat{D}$ excludes infinite pencil eigenvalues.
	These restrictions on the pencil parameter allow zero-energy states satisfying $\hat{A}\lvert\psi_{0}\rangle=0$.
	Normal- and antinormal-ordered quadratic products 
	\begin{equation}
		\begin{aligned}
			\hat{\mathcal{H}}_{-} =&~ \hat{A}^{\dagger}\hat{A} = \hat{a}^{\dagger}\hat{a}\hat{D}^{\dagger}\hat{D} + \hat{a}^{\dagger}\hat{D}^{\dagger}\hat{C} + \hat{a}\hat{C}^{\dagger}\hat{D} + \hat{C}^{\dagger}\hat{C}, \\
			\hat{\mathcal{H}}_{+} =&~ \hat{A}\hat{A}^{\dagger} = \left( \hat{a}^{\dagger}\hat{a} + 1 \right)\hat{D}\hat{D}^{\dagger} + \hat{a}\hat{D}\hat{C}^{\dagger} + \hat{a}^{\dagger}\hat{C}\hat{D}^{\dagger} + \hat{C}\hat{C}^{\dagger},
		\end{aligned}
		\label{pair}
	\end{equation}
	define two dimensionless factorized Hamiltonians.
	The products $\hat{C}^{\dagger}\hat{C}$ and $\hat{D}^{\dagger}\hat{D}$ in $\hat{\mathcal{H}}_{-}$, and $\hat{C}\hat{C}^{\dagger}$ and $\hat{D}\hat{D}^{\dagger}$ in $\hat{\mathcal{H}}_{+}$, give static and number-dependent spin terms with identity, longitudinal, and transverse components.
	The mixed products multiply $\hat{a}$ or $\hat{a}^{\dagger}$ and generate spin-independent and longitudinal drives and rotating and counter-rotating couplings.
	
	We use the normal product to define our physical spin-pencil Hamiltonian after choosing one positive frequency scale $\Omega_{\mathrm{F}}$,
	\begin{widetext}
		\begin{align}
			\begin{aligned}
				\frac{\hat{H}_{\mathrm{F}}}{\hbar} = \Omega_{\mathrm{F}}\hat{\mathcal{H}}_{-} =&~ \omega\hat{a}^{\dagger}\hat{a} + \frac{1}{2}\omega_{0}\hat{\sigma}_{z} + g_{1}\left( \hat{Q}_{+} + \hat{Q}_{-} \right) + g_{2}\left( \hat{R}_{+} + \hat{R}_{-} \right) + \chi\left( \hat{a}^{\dagger}\hat{a} + \frac{1}{2} \right)\hat{\sigma}_{z} \\
				&+ \epsilon\left( \hat{a}^{\dagger} + \hat{a} \right)  + \nu \left( \hat{\sigma}_{+} + \hat{\sigma}_{-} \right) + \lambda\left( \hat{a}^{\dagger} + \hat{a} \right)\hat{\sigma}_{z} + \kappa\hat{a}^{\dagger}\hat{a}\left( \hat{\sigma}_{+} + \hat{\sigma}_{-} \right) + \varepsilon_{\mathrm{F}}.
			\end{aligned}
		\end{align}
	\end{widetext}
	It contains anisotropic Rabi terms~\cite{Xie2014p021046,Wang2019p4569,Chen2021p043708,Kafuri2024pC82}, a Rabi--Stark deformation~\cite{Eckle2017p294004,Xie2019p245304,Braak2024pC97}, direct bosonic and spin drives~\cite{Henriet2014p023820,CastanosCervantes2021p033709}, a longitudinal spin-dependent bosonic drive~\cite{Richer2016p134501,Potts2025p153603}, and an intensity-dependent transverse spin drive~\cite{Buck1981p132,RodriguezLara2013p12888,RodriguezLara2014p1719,ArmentaRico2020p063825,Duan2022p083045}, with frequencies
	\begin{widetext}
		\begin{align}
			\begin{aligned}
				\omega =&~ \Omega_{\mathrm{F}}\left[ d_{0}^{2} + \frac{1}{2}\left( d_{+}^{2} + d_{-}^{2} \right) \right], \qquad &
				\omega_{0} =&~ \Omega_{\mathrm{F}}\left[ c_{-}^{2} - c_{+}^{2} - \frac{1}{2}\left( d_{-}^{2} - d_{+}^{2} \right) \right], \\
				g_{1} =&~ \Omega_{\mathrm{F}}\left( c_{0}d_{+} + c_{-}d_{0} \right), \qquad &
				g_{2} =&~ \Omega_{\mathrm{F}}\left( c_{0}d_{-} + c_{+}d_{0} \right), \\
				\chi =&~ \frac{1}{2}\Omega_{\mathrm{F}}\left( d_{-}^{2} - d_{+}^{2} \right), \qquad &
				\epsilon =&~ \Omega_{\mathrm{F}}\left[ c_{0}d_{0} + \frac{1}{2}\left( c_{-}d_{-} + c_{+}d_{+} \right) \right], \\
				\nu =&~ \Omega_{\mathrm{F}}c_{0}\left( c_{+} + c_{-} \right), \qquad &
				\lambda =&~ \frac{1}{2}\Omega_{\mathrm{F}}\left( c_{-}d_{-} - c_{+}d_{+} \right), \\
				\kappa =&~ \Omega_{\mathrm{F}}d_{0}\left( d_{+} + d_{-} \right), \qquad &
				\varepsilon_{\mathrm{F}} =&~ \Omega_{\mathrm{F}}\left[ c_{0}^{2} + \frac{1}{2}\left( c_{+}^{2} + c_{-}^{2} \right) \right].
			\end{aligned}
		\end{align}
	\end{widetext}
	The six real pencil parameters determine the nine Hamiltonian frequencies and the energy offset $\varepsilon_{\mathrm{F}}$ up to the common scale $\Omega_{\mathrm{F}}$, constraining the coupling ratios to a codimension-four spin-pencil submanifold.
	For $\Omega_{\mathrm{F}}>0$, the factorization gives $\langle\psi\rvert\hat{H}_{\mathrm{F}}\lvert\psi\rangle=\hbar\Omega_{\mathrm{F}}\lVert\hat{A}\lvert\psi\rangle\rVert^{2}\geq0$, including in the presence of the boson-dependent transverse spin drive $\kappa\hat{a}^{\dagger}\hat{a}\left( \hat{\sigma}_{+} + \hat{\sigma}_{-} \right)$ and the parity-breaking transverse bias $\nu\left( \hat{\sigma}_{+} + \hat{\sigma}_{-} \right)$.
	
	\section{Zero modes}
	\label{sec:zmodes}
	Our factorized Hamiltonians obey the intertwining relations
	\begin{align}
		\hat{A}\hat{\mathcal{H}}_{-} =&~ \hat{\mathcal{H}}_{+}\hat{A}, \qquad &
		\hat{A}^{\dagger}\hat{\mathcal{H}}_{+} =&~ \hat{\mathcal{H}}_{-}\hat{A}^{\dagger},
	\end{align}
	which pair their nonzero spectra. For a normalized state with positive energy,
	\begin{align}
		\hat{\mathcal{H}}_{-}\lvert \psi \rangle =&~ E\lvert \psi \rangle, \qquad
		E > 0, \qquad
		\langle \psi\vert\psi\rangle = 1,
	\end{align}
	the image under $\hat{A}$ has squared norm
	\begin{align}
		\left\Vert \hat{A}\lvert \psi \rangle \right\Vert^{2} =&~ E,
	\end{align}
	and retains the same energy,
	\begin{align}
		\hat{\mathcal{H}}_{+} \hat{A} \lvert \psi \rangle =&~ \hat{A} \hat{\mathcal{H}}_{-}\lvert \psi \rangle = E \hat{A} \lvert \psi \rangle.
	\end{align}
	The operator $\hat{A}/\sqrt{E}$ gives an isometric bijection between the $E$-eigenspaces of $\hat{\mathcal{H}}_{-}$ and $\hat{\mathcal{H}}_{+}$, with inverse $\hat{A}^{\dagger}/\sqrt{E}$,
	\begin{align}
		\mathrm{spec}\,\hat{\mathcal{H}}_{-}\setminus\{0\} =&~
		\mathrm{spec}\,\hat{\mathcal{H}}_{+}\setminus\{0\},
	\end{align}
	with equal multiplicities.
	Only the zero modes of $\hat{\mathcal{H}}_{\pm}$ can remain unpaired.

	\subsection{\texorpdfstring{Zero modes of $\hat{\mathcal{H}}_-$}{Zero modes of H-}}
	\label{sec:zmodes-}
	We obtain the zero modes of $\hat{\mathcal{H}}_-$ from the factor equation
	\begin{align}
		\hat{A}\lvert \psi_{0} \rangle =&~
		\left( \hat{C} + \hat{a}\hat{D} \right)\lvert \psi_{0} \rangle = 0.
	\end{align}
	Our regular pencil factors through the annihilation operator,
	\begin{align}
		\hat{A} =&~ \hat{D}\left( \hat{a}\hat{I}_{2} + \hat{M} \right), \qquad
		\hat{M} = \hat{D}^{-1}\hat{C},
	\end{align}
	so the invertibility of $\hat{D}$ puts the zero-mode sector in standard form,
	\begin{align}
		\ker\hat{A} =&~ \ker\left( \hat{a}\hat{I}_{2} + \hat{M} \right).
	\end{align}
	Let $\lvert\alpha\rangle$ be the bosonic coherent state satisfying $\hat{a}\lvert\alpha\rangle=\alpha\lvert\alpha\rangle$.
	For the product $\lvert \psi_{0} \rangle = \lvert \alpha \rangle \otimes \lvert m_{\pm} \rangle$, with spin eigenvector
	\begin{align}
		\hat{M}\lvert m_{\pm} \rangle =&~ \mu_{\pm}\lvert m_{\pm} \rangle,
	\end{align}
	the factor equation requires the coherent amplitude to cancel the spin eigenvalue,
	\begin{align}
		\left( \alpha + \mu_{\pm} \right)\lvert \alpha \rangle \otimes \lvert m_{\pm} \rangle =&~ 0.
	\end{align}
	The coherent displacement $\alpha = -\mu_{\pm}$ follows from the two eigenvalues of the standard matrix,
	\begin{align}
		\mu_{\pm} =&~
		\frac{ 2d_{0}c_{0} - d_{+}c_{-} - d_{-}c_{+} \pm \sqrt{\Delta_{\mathrm{M}}} }
		{ 2\left( d_{0}^{2} - d_{+}d_{-} \right) },
	\end{align}
	and discriminant
	\begin{align}
		\Delta_{\mathrm{M}} =&~
		\left( 2d_{0}c_{0} - d_{+}c_{-} - d_{-}c_{+} \right)^{2}
		- 4\left( d_{0}^{2} - d_{+}d_{-} \right)
		\left( c_{0}^{2} - c_{+}c_{-} \right).
	\end{align}
	These displacements solve the pencil-root condition,
	\begin{align}
		\det\hat{P}(\alpha) =&~
		\det\hat{D}\,
		\det\left( \hat{M} + \alpha\hat{I}_{2} \right) = 0,
	\end{align}
	so the coherent zero modes recover the generalized eigenvalue problem encoded by the spin pencil.
	
	The discriminant gives the regular-pencil trichotomy for the coherent displacements.
	For $\Delta_{\mathrm{M}}>0$, the pencil has two distinct real roots and our zero-mode sector has two real coherent displacements.
	For $\Delta_{\mathrm{M}}<0$, the pencil has a complex-conjugate pair and our zero modes have complex-conjugate coherent displacements.
	For $\Delta_{\mathrm{M}}=0$, the pencil has a repeated root, which is defective when $\hat{M}$ has only one independent eigenvector and diagonalizable when $\hat{M}$ is scalar.
	
	For $\Delta_{\mathrm{M}}\neq0$, we choose the normalized spin eigenvectors
	\begin{align}
		\begin{aligned}
			\lvert m_{\pm} \rangle =&~ \frac{1}{\mathcal{N}_{\pm}}
			\begin{pmatrix}
				d_{0}c_{+} - d_{+}c_{0} \\
				\mu_{\pm}\left( d_{0}^{2} - d_{+}d_{-} \right) - d_{0}c_{0} + d_{+}c_{-}
			\end{pmatrix}, \\
			\mathcal{N}_{\pm}^{2} =&~ \lvert d_{0}c_{+} - d_{+}c_{0} \rvert^{2} + \lvert \mu_{\pm}\left( d_{0}^{2} - d_{+}d_{-} \right) - d_{0}c_{0} + d_{+}c_{-} \rvert^{2}.
		\end{aligned}
	\end{align}
	If this eigenvector vanishes at isolated parameter values, we use the equivalent choice
	\begin{align}
		\begin{aligned}
			\lvert m_{\pm} \rangle =&~ \frac{1}{\widetilde{\mathcal{N}}_{\pm}}
			\begin{pmatrix}
				\mu_{\pm}\left( d_{0}^{2} - d_{+}d_{-} \right) - d_{0}c_{0} + d_{-}c_{+} \\
				d_{0}c_{-} - d_{-}c_{0}
			\end{pmatrix}, \\
			\widetilde{\mathcal{N}}_{\pm}^{2} =&~ \lvert d_{0}c_{-} - d_{-}c_{0} \rvert^{2} + \lvert \mu_{\pm}\left( d_{0}^{2} - d_{+}d_{-} \right) - d_{0}c_{0} + d_{-}c_{+} \rvert^{2}.
		\end{aligned}
	\end{align}
	The zero-mode equation gives two coherent spin modes,
	\begin{align}
		\lvert \psi_{0}^{(\pm)} \rangle =&~ \left\vert -\mu_{\pm} \right\rangle \otimes \lvert m_{\pm} \rangle.
	\end{align}
	
	For $\Delta_{\mathrm{M}}<0$, the amplitudes $\mu_{\pm}$ form a complex-conjugate pair and the two normalizable coherent spin modes are related by complex conjugation.
	
	At $\Delta_{\mathrm{M}}=0$, the standard-form spin matrix is diagonalizable only when $\hat{C}=\mu\hat{D}$.
	Otherwise, $\hat{M}$ has a Jordan spin chain,
	\begin{align}
		\left( \hat{M} - \mu\hat{I}_{2} \right)\lvert m_{0} \rangle =&~ 0, \qquad
		\left( \hat{M} - \mu\hat{I}_{2} \right)\lvert m_{1} \rangle = \lvert m_{0} \rangle.
	\end{align}
	The defective-pencil zero-mode doublet follows from a Fock expansion,
	\begin{align}
		\lvert \psi_{0}^{(j)} \rangle =&~ \sum_{n = 0}^{\infty} \lvert n \rangle \otimes \frac{\left( -\hat{M} \right)^{n}}{\sqrt{n!}}\lvert m_{j} \rangle, \qquad
		j \in \left\{ 0,1 \right\}.
	\end{align}
	The coherent product modes cover the diagonalizable pencil sectors.
	The full Fock recursion shows that the kernel dimension does not depend on diagonalizability.
	The Fock expansion
	\begin{align}
		\lvert \psi_{0} \rangle =&~ \sum_{n = 0}^{\infty} \lvert n \rangle \otimes \lvert w_{n} \rangle
	\end{align}
	turns the zero-mode equation into the coefficient recursion
	\begin{align}
		\sqrt{n + 1} \, \lvert w_{n + 1} \rangle + \hat{M}\lvert w_{n} \rangle =&~ 0,
	\end{align}
	whose iteration determines each coefficient from the initial spin vector,
	\begin{align}
		\lvert w_{n} \rangle =&~ \frac{\left( -\hat{M} \right)^{n}}{\sqrt{n!}}\lvert w_{0} \rangle.
	\end{align}
	The norm bound
	\begin{align}
		\lVert w_{n} \rVert \leq&~ \frac{\lVert \hat{M} \rVert^{n}}{\sqrt{n!}}\lVert w_{0} \rVert
	\end{align}
	makes the series normalizable for every spin vector $\lvert w_{0} \rangle$.
	The two free components of $\lvert w_{0} \rangle$ give
	\begin{align}
		\dim\ker\hat{A} =&~ 2.
	\end{align}
	
	\subsection{\texorpdfstring{Zero modes of $\hat{\mathcal{H}}_+$}{Zero modes of H+}}
	\label{sec:zmodes+}
	The adjoint zero-mode equation,
	\begin{align}
		\hat{A}^{\dagger}\lvert \psi \rangle =&~ 0, \qquad
		\lvert \psi \rangle = \sum_{n = 0}^{\infty} \lvert n \rangle \otimes \lvert v_{n} \rangle,
	\end{align}
	has level-zero condition
	\begin{align}
		\hat{C}^{\dagger}\lvert v_{0} \rangle =&~ 0.
	\end{align}
	The invertibility of $\hat{C}$ forces $\lvert v_{0} \rangle = 0$, and the recursion sets $\lvert v_{n} \rangle = 0$ for all $n$.
	The adjoint factor has a trivial kernel,
	\begin{align}
		\dim\ker\hat{A}^{\dagger} =&~ 0.
	\end{align}
	The two recursions establish the kernel dimensions throughout the regular spin-pencil class,
	\begin{align}
		\dim\ker\hat{A} =&~ 2, &
		\dim\ker\hat{A}^{\dagger} =&~ 0,
	\end{align}
	whose difference determines the factor index
	\begin{align}
		\mathrm{ind}\,\hat{A} =&~ \dim\ker\hat{A}-\dim\ker\hat{A}^{\dagger}=2.
	\end{align}
	The factor index coincides with the Witten index of the auxiliary Hamiltonian $\operatorname{diag}(\hat{\mathcal{H}}_{-},\hat{\mathcal{H}}_{+})$, graded by $\operatorname{diag}(\hat{I},-\hat{I})$.
	This grading acts on two copies of the boson--spin Hilbert space and is distinct from the physical spin grading.
	The normal product $\hat{\mathcal{H}}_{-}=\hat{A}^{\dagger}\hat{A}$ is nonnegative throughout the class and has an exact doubly degenerate zero-eigenvalue ground multiplet.
	
	\section{Bargmann coherent-gauge hierarchy}
	\label{sec:Sec4}
	
	In the distinct-root sector of our regular spin-pencil class, $\Delta_{\mathrm{M}} \neq 0$, we construct the exact coherent spin zero modes by displacing the bosonic vacuum,
	\begin{align}
		\lvert \psi_{0}^{(\pm)} \rangle =&~  \left\vert -\mu_{\pm} \right\rangle \otimes \lvert m_{\pm} \rangle
		= \hat{D}_{\mathrm{b}}\left( -\mu_{\pm} \right)\lvert 0 \rangle \otimes \lvert m_{\pm} \rangle,
	\end{align}
	with bosonic displacement operator $\hat{D}_{\mathrm{b}}(\alpha) = e^{ \alpha \hat{a}^{\dagger} - \alpha^{\ast} \hat{a} }$.
	They solve the zero-mode equation $\hat{A}\lvert \psi_{0}^{(\pm)} \rangle = 0$ and give exact zero-energy eigenstates,
	\begin{align}
		\hat{H}_{\mathrm{F}}\lvert \psi_{0}^{(\pm)} \rangle =&~ \hbar\Omega_{\mathrm{F}}\hat{A}^{\dagger}\hat{A}\lvert \psi_{0}^{(\pm)} \rangle = 0.
	\end{align}
	
	We use the analytic Bargmann dual $\langle z\rvert=\sum_{n=0}^{\infty}z^{n}\langle n\rvert/\sqrt{n!}$.
	The Bargmann substitutions $\hat{a}^{\dagger} \to z$ and $\hat{a} \to d/dz$ map the coherent zero modes to normalized holomorphic wavefunctions,
	\begin{align}
		\psi_{0}^{(\pm)}(z) \equiv \langle z \vert \psi_{0}^{(\pm)} \rangle =&~ e^{-\frac{1}{2}\lvert\mu_{\pm}\rvert^{2}} e^{-\mu_{\pm}z}\lvert m_{\pm} \rangle,
	\end{align}
	where $e^{-\frac{1}{2}\lvert\mu_{\pm}\rvert^{2}}$ provides the coherent-state normalization.
	We call $e^{-\mu_{\pm}z}$ a coherent gauge because removing this scalar exponential from the wavefunction shifts the Bargmann derivative by $-\mu_{\pm}$.
	Under these substitutions, the dimensionless $\hat{\mathcal{H}}_{-}$ acts as a first-order Bargmann differential operator,
	\begin{align}
		\hat{\mathcal{H}}_{-} =&~
		\hat{D}^{\dagger}\hat{D}\, z \frac{d}{dz}
		+ \hat{D}^{\dagger}\hat{C}\, z
		+ \hat{C}^{\dagger}\hat{D}\, \frac{d}{dz}
		+ \hat{C}^{\dagger}\hat{C}.
	\end{align}
	
	\subsection{Baselines}
	\label{sec:baselines}
	We use our exact coherent zero modes as gauges to build a rung-by-rung hierarchy of baseline energies.
	The baselines give reference energies for a finite-polynomial problem and become exact eigenenergies at exceptional parameter values where the Bargmann spinor truncates~\cite{Judd1979p1685,Kus1985p2792,Moroz2012p60010,Moroz2013p319,Moroz2014p252,Moroz2014p495204,Moroz2016p50004,Moroz2018p295201,Kimoto2021p9458,Chen2021p043708}.
	We keep the scalar gauge independent of the spinor coefficient and allow the spinor coefficient to have finite degree,
	\begin{align}
		\psi_{N}^{(\mu_{\pm})}(z) =&~ e^{-\mu_{\pm} z} \sum_{k = 0}^{N} z^{k} \lvert \phi_{k}^{(N,\mu_{\pm})} \rangle, \qquad
		\lvert \phi_{k}^{(N,\mu_{\pm})} \rangle \in \mathbb{C}^{2}, \qquad
		\lvert \phi_{N}^{(N,\mu_{\pm})} \rangle \neq 0,
	\end{align}
	where $N=0$ recovers the unnormalized zero modes with $\lvert \phi_{0}^{(0,\mu_{\pm})} \rangle=\lvert m_{\pm} \rangle$.
	Factoring out the scalar exponential conjugates the Bargmann Hamiltonian and shifts $d/dz \to d/dz-\mu_{\pm}$,
	\begin{align}
		\begin{aligned}
			\hat{h}_{\mu_{\pm}} =&~ e^{\mu_{\pm} z}\hat{\mathcal{H}}_{-}e^{-\mu_{\pm} z} \\
			=&~ \hat{D}^{\dagger}\hat{D}\left( \hat{M} - \mu_{\pm}\hat{I}_{2} \right) z
			+ \hat{D}^{\dagger}\hat{D}\, z\frac{d}{dz}
			\\+&~ \hat{C}^{\dagger}\hat{D}\left( \hat{M} - \mu_{\pm}\hat{I}_{2} \right)
			+ \hat{C}^{\dagger}\hat{D}\, \frac{d}{dz},
		\end{aligned}
	\end{align}
	so that the zero-mode spinors satisfy
	\begin{align}
		\hat{h}_{\mu_{\pm}}\lvert m_{\pm} \rangle =&~ 0.
	\end{align}
	Any higher-rung solution with $N \geq 1$ must cancel the degree-$(N+1)$ contribution produced by the top spinor coefficient $\lvert \phi_{N}^{(N,\mu_{\pm})} \rangle$.
	Since $\hat{D}$ is invertible, this cancellation reduces to
	\begin{align}
		\hat{M}\lvert \phi_{N}^{(N,\mu_{\pm})} \rangle =&~ \mu_{\pm}\lvert \phi_{N}^{(N,\mu_{\pm})} \rangle,
	\end{align}
	and places the top polynomial coefficient in the same spin eigenspace as the zero mode,
	\begin{align}
		\lvert \phi_{N}^{(N,\mu_{\pm})} \rangle =&~ c_{\pm}\lvert m_{\pm} \rangle, \qquad
		c_{\pm} \neq 0.
	\end{align}
	
	To extract the baseline energies, we look for finite-polynomial solutions
	\begin{align}
		\hat{h}_{\mu_{\pm}}\sum_{k = 0}^{N} z^{k}\lvert \phi_{k}^{(N,\mu_{\pm})} \rangle =&~
		E_{N}^{(\mu_{\pm})} \sum_{k = 0}^{N} z^{k}\lvert \phi_{k}^{(N,\mu_{\pm})} \rangle.
	\end{align}
	Matching the coefficient of $z^{N}$ couples the top two spinor coefficients through
	\begin{align}
		\begin{aligned}
			& \left[ N \hat{D}^{\dagger}\hat{D} + \hat{C}^{\dagger}\hat{D}\left( \hat{M} - \mu_{\pm}\hat{I}_{2} \right) - E_{N}^{(\mu_{\pm})} \right]\lvert \phi_{N}^{(N,\mu_{\pm})} \rangle
			=&~- \hat{D}^{\dagger}\hat{D}\left( \hat{M} - \mu_{\pm}\hat{I}_{2} \right)\lvert \phi_{N-1}^{(N,\mu_{\pm})} \rangle.
		\end{aligned}
	\end{align}
	Since $\hat{M}$ need not be Hermitian, we introduce its left spin eigenvectors,
	\begin{align}
		\langle n_{\pm} \rvert \hat{M} =&~ \mu_{\pm}\langle n_{\pm} \rvert.
	\end{align}
	For distinct roots, $(\mu_i-\mu_j)\langle n_i\vert m_j\rangle=0$ for $i,j\in\{+,-\}$, so opposite-branch left and right eigenvectors are biorthogonal, while the right eigenvectors need not be mutually orthogonal.
	Multiplying the level-$z^{N}$ equation by $\langle n_{\pm}\rvert\left( \hat{D}^{\dagger}\hat{D} \right)^{-1}$ from the left removes the lower coefficient because $\langle n_{\pm}\rvert(\hat{M}-\mu_{\pm}\hat{I}_{2})=0$.
	The rightmost factor $(\hat{M}-\mu_{\pm}\hat{I}_{2})$ also annihilates $\lvert\phi_N^{(N,\mu_{\pm})}\rangle$ in the middle term, giving
	\begin{align}
		\begin{aligned}
			\langle n_{\pm}\rvert\left( \hat{D}^{\dagger}\hat{D} \right)^{-1}
			\left[ N \hat{D}^{\dagger}\hat{D} + \hat{C}^{\dagger}\hat{D}\left( \hat{M} - \mu_{\pm}\hat{I}_{2} \right)
			- E_{N}^{(\mu_{\pm})} \right]\lvert \phi_{N}^{(N,\mu_{\pm})} \rangle =&~ 0, \\
			N \langle n_{\pm}\vert m_{\pm} \rangle
			- E_{N}^{(\mu_{\pm})} \langle n_{\pm}\rvert\left( \hat{D}^{\dagger}\hat{D} \right)^{-1}\lvert m_{\pm} \rangle =&~ 0.
		\end{aligned}
	\end{align}
	For distinct roots, the same-branch overlap $\langle n_{\pm}\vert m_{\pm} \rangle$ is nonzero.
	Provided $\langle n_{\pm}\rvert\left( \hat{D}^{\dagger}\hat{D} \right)^{-1}\lvert m_{\pm} \rangle \neq 0$, the projected equation defines the finite baseline energy
	\begin{align}
		E_{N}^{(\mu_{\pm})} =&~ N \gamma_{\pm},
	\end{align}
	with branch slopes
	\begin{align}
		\frac{1}{\gamma_{\pm}} =&~
		\frac{\langle n_{\pm}\rvert\left( \hat{D}^{\dagger}\hat{D} \right)^{-1}\lvert m_{\pm} \rangle}{\langle n_{\pm}\vert m_{\pm} \rangle}
		=
		\operatorname{Tr} \left[ \hat{P}_{\pm}\left( \hat{D}^{\dagger}\hat{D} \right)^{-1} \right],
	\end{align}
	where we use the biorthogonal spectral projectors
	\begin{align}
		\hat{P}_{\pm} =&~ \frac{\lvert m_{\pm} \rangle\langle n_{\pm} \rvert}{\langle n_{\pm}\vert m_{\pm} \rangle},
		\qquad
		\hat{P}_{+} + \hat{P}_{-} = \hat{I}_{2}.
	\end{align}
	The positivity of $\hat{D}^{\dagger}\hat{D}$ does not guarantee this nonvanishing condition because the left and right spin eigenvectors need not be adjoints.
	If the numerator of the reciprocal-slope formula vanishes, the projected equation forces $N=0$, so no polynomial solution of degree $N\geq1$ exists in that gauge and its coherent zero mode remains.
	When both branch slopes are finite, they satisfy
	\begin{align}
		\frac{1}{\gamma_{+}} + \frac{1}{\gamma_{-}} =&~ \operatorname{Tr} \left( \hat{D}^{\dagger}\hat{D} \right)^{-1}.
	\end{align}
	
	For the branch $\mu_{\pm}$ and baseline energy $E_{N}^{(\mu_{\pm})}$, matching the coefficient of $z^{k}$ gives the three-term recurrence
	\begin{align}
		\begin{aligned}
			\left( k + 1 \right)\hat{C}^{\dagger}\hat{D}\,\lvert \phi_{k+1}^{(N,\mu_{\pm})} \rangle =&~
			-\left[ k \hat{D}^{\dagger}\hat{D} + \hat{C}^{\dagger}\hat{D}\left( \hat{M} - \mu_{\pm}\hat{I}_{2} \right) - E_{N}^{(\mu_{\pm})} \right] \lvert \phi_{k}^{(N,\mu_{\pm})} \rangle \\
			&~ - \hat{D}^{\dagger}\hat{D}\left( \hat{M} - \mu_{\pm}\hat{I}_{2} \right)\lvert \phi_{k-1}^{(N,\mu_{\pm})} \rangle,
		\end{aligned}
	\end{align}
	with endpoint conditions
	\begin{align}
		\lvert \phi_{-1}^{(N,\mu_{\pm})} \rangle =&~ 0, \quad
		\lvert \phi_{N}^{(N,\mu_{\pm})} \rangle = c_{\pm}\lvert m_{\pm} \rangle, \quad
		c_{\pm} \neq 0, \quad
		\lvert \phi_{N+1}^{(N,\mu_{\pm})} \rangle = 0.
	\end{align}
	The degree-$(N+1)$ condition has rank one for $\Delta_{\mathrm{M}} \neq 0$ because
	$\hat{D}^{\dagger}\hat{D}\left( \hat{M} - \mu_{\pm}\hat{I}_{2} \right)$
	has a one-dimensional range, while the projected level-$z^{N}$ equation fixes the baseline energy.
	After we remove the overall spinor scale, the remaining endpoint problem leaves one scalar compatibility condition for each rung and branch.
	This scalar compatibility condition defines the rung-$N$ constraint,
	\begin{align}
		P_{N}^{(\mu_{\pm})} =&~ 0, \qquad N \geq 1.
	\end{align}
	The rung $N=0$ is constraint free and has $\lvert \phi_{0}^{(0,\mu_{\pm})} \rangle = \lvert m_{\pm} \rangle$, reproducing the zero-mode doublet.
	These endpoint compatibility conditions extend the constraint-polynomial construction of isolated Rabi eigenstates~\cite{Kus1985p2792,Moroz2018p295201,Kimoto2021p9458,Chen2021p043708} to our regular spin-pencil class.
	
	For $\Delta_{\mathrm{M}} < 0$, the same diagonalizable construction applies over the complex spin space.
	Since $\hat{M}$ is real, the two gauges and spin eigenvectors occur as conjugate pairs,
	\begin{align}
		\mu_{-} =&~ \mu_{+}^{\ast}, \qquad
		\lvert m_{-} \rangle = \lvert m_{+} \rangle^{\ast}, \qquad
		\langle n_{-} \rvert = \langle n_{+} \rvert^{\ast}.
	\end{align}
	The coherent zero modes remain normalizable Bargmann states, with conjugate Taylor coefficients in the holomorphic basis,
	\begin{align}
		\psi_{0}^{(-)}(z) \equiv \langle z \vert \psi_{0}^{(-)} \rangle =&~ e^{-\frac{1}{2}\lvert\mu_{+}\rvert^{2}} e^{-\mu_{+}^{\ast}z}\lvert m_{+} \rangle^{\ast},
	\end{align}
	up to the corresponding spin normalization.
	The branch baselines form a conjugate pair,
	\begin{align}
		E_{N}^{(\mu_{-})} =&~ \left[ E_{N}^{(\mu_{+})} \right]^{\ast},
	\end{align}
	so the Hermiticity of $\hat{H}_{\mathrm{F}}$ allows an excited finite-degree rung only when the corresponding baseline is real and the endpoint compatibility condition is satisfied.
	The zero-mode doublet is unaffected by this restriction because $E_{0}^{(\mu_{\pm})}=0$.
	
	At the discriminant surface $\Delta_{\mathrm{M}} = 0$, the two coherent gauges coalesce,
	\begin{align}
		\mu_{+} =&~ \mu_{-} = \mu_{0},
	\end{align}
	and the diagonalizable branch construction no longer applies as two independent sectors.
	We separate the coalesced eigenvalue from the nilpotent part of the real invertible spin matrix,
	\begin{align}
		\hat{M} =&~ \mu_{0}\hat{I}_{2} + \hat{J}, \qquad
		\hat{J} = \hat{M} - \mu_{0}\hat{I}_{2}, \qquad
		\hat{J}^{2} = 0,
	\end{align}
	with $\hat{J}=0$ for the scalar member and $\hat{J}\neq0$ for the defective Jordan member.
	The Bargmann zero-mode equation reduces to a constant-matrix differential system,
	\begin{align}
		\left( \frac{d}{dz}\hat{I}_{2} + \hat{M} \right)\psi_{0}(z) =&~ 0,
	\end{align}
	with matrix-exponential solution
	\begin{align}
		\psi_{0}(z) =&~ e^{-\mu_{0}z}e^{-\hat{J}z}\lvert \chi \rangle
		= e^{-\mu_{0}z}\left( \hat{I}_{2} - z\hat{J} \right)\lvert \chi \rangle.
	\end{align}
	For the defective member, a Jordan chain
	\begin{align}
		\hat{J}\lvert m_{0} \rangle =&~ 0, \qquad
		\hat{J}\lvert \eta_{0} \rangle = \lvert m_{0} \rangle,
	\end{align}
	gives two independent zero modes,
	\begin{align}
		\begin{aligned}
			\psi_{0,1}^{(\mu_{0})}(z) =&~ e^{-\mu_{0}z}\lvert m_{0} \rangle, \\
			\psi_{0,2}^{(\mu_{0})}(z) =&~ e^{-\mu_{0}z}\left( \lvert \eta_{0} \rangle - z\lvert m_{0} \rangle \right).
		\end{aligned}
	\end{align}
	The coalesced sector retains a two-dimensional zero-mode kernel, but one zero mode becomes a finite-degree Jordan-dressed coherent state instead of a second independent coherent-gauge branch.
	At the coalescence, the two-branch spectral resolution ceases to define independent branch projectors; in the defective limit the branch projectors become singular, while in the scalar member their separation is nonunique.
	The slope formula based on the distinct-root spectral resolution no longer applies, so the finite-degree hierarchy must be rebuilt with the single gauge $\mu_{0}$ and the corresponding scalar or Jordan spin structure.
	
	\section{Discussion}
	\label{sec:Sec5}
	
	We classify the spin matrices by the identity line and exchange plane,
	\begin{align}
		\mathcal{V}_{0} =&~ \operatorname{span}_{\mathbb{R}}\{\hat{I}_{2}\}, \qquad
		\mathcal{V}_{1} = \operatorname{span}_{\mathbb{R}}\{\hat{\sigma}_{+},\hat{\sigma}_{-}\}.
	\end{align}
	Conjugation by $\hat{\sigma}_{z}$ leaves $\mathcal{V}_{0}$ fixed and reverses the sign of $\mathcal{V}_{1}$, making them the even and odd spin-operator subspaces.
	The spin-inversion charge $\hat{B}=\hat{\sigma}_{z}/2$ satisfies $[\hat{B},\hat{I}_{2}]=0$ and $[\hat{B},\hat{\sigma}_{\pm}]=\pm\hat{\sigma}_{\pm}$, assigning charges $0,+1,-1$ to the basis operators while the coefficients $\xi_j$ remain scalars.
	
	Independent choices of subspace for $\hat{C}$ and $\hat{D}$ give the sector square in Table~\ref{tab:sector-square}, which relates the four limiting pencils to their physical models.
	The identity--identity choice, $(\hat{C},\hat{D}) = (c_{0}\hat{I}_{2},d_{0}\hat{I}_{2})$, reduces to a driven bosonic oscillator with a spectator spin.
	Choosing the exchange plane for $\hat{C}$ while retaining $\hat{D}$ on the identity line, $(\hat{C},\hat{D}) = (c_{-}\hat{\sigma}_{-} + c_{+}\hat{\sigma}_{+},d_{0}\hat{I}_{2})$, gives the Tomka--Pletyukhov--Gritsev supersymmetric factorization sheet~\cite{Tomka2015p13097}.
	The complementary mixed choice, $(\hat{C},\hat{D}) = (c_{0}\hat{I}_{2},d_{-}\hat{\sigma}_{-} + d_{+}\hat{\sigma}_{+})$, gives our anisotropic Rabi--Stark factorization sheet~\cite{Kafuri2024pC82}, while placing both matrices in the exchange plane, $(\hat{C},\hat{D}) = (c_{-}\hat{\sigma}_{-} + c_{+}\hat{\sigma}_{+},d_{-}\hat{\sigma}_{-} + d_{+}\hat{\sigma}_{+})$, gives two driven bosonic oscillators conditioned by the spin-inversion charge, with $\hat{M}$ diagonal in the $\hat{\sigma}_{z}$ basis.
	\begin{table}[t]
		\caption{Sector square for the two spin matrices. The II and EE corners have complete displaced-oscillator spectra. In the mixed corners, excited finite-polynomial states require endpoint compatibility in general.}
		\label{tab:sector-square}
		\begin{ruledtabular}
			\begin{tabular}{lcc}
				& $\hat{C}\in\mathcal{V}_{0}$ & $\hat{C}\in\mathcal{V}_{1}$ \\
				\hline
				\noalign{\vskip 0.8ex}
				$\hat{D}\in\mathcal{V}_{0}$ & \shortstack{II\\Driven oscillator} & \shortstack{EI\\Anisotropic Rabi sheet} \\[1ex]
				$\hat{D}\in\mathcal{V}_{1}$ & \shortstack{IE\\Anisotropic Rabi--Stark sheet} & \shortstack{EE\\Spin-conditioned oscillators} \\[0.6ex]
			\end{tabular}
		\end{ruledtabular}
	\end{table}
	
	The generic pencil has both identity and exchange components in each matrix and includes the remaining couplings.
	Within each corner we label the two branches by their spin eigenvector, since these labels need not agree with those assigned by the sign of $\sqrt{\Delta_{\mathrm{M}}}$, as happens at the corners with $\det\hat{D}<0$.
	
	\subsection{Identity--identity pencil for a decoupled driven bosonic mode}
	\label{sec:Sec51}
	
\begin{figure}[t]
		\centering
		\includegraphics[scale=1]{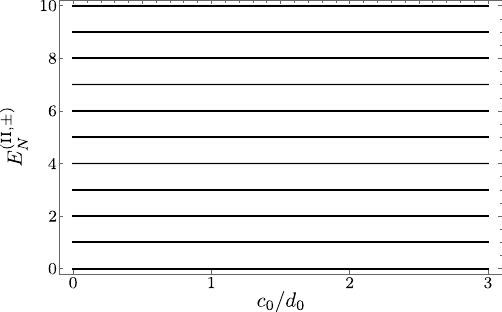}
		\caption{
			Identity--identity spectral branch.
			Light gray curves show the numerical finite-Fock spectrum of $\hat{\mathcal{H}}_{\mathrm{II}-}$ for cutoff $N_{\mathrm{F}}=1000$, and solid black curves show the coherent-gauge baselines $E_{N}^{(\mathrm{II},\pm)}=Nd_{0}^{2}$.
			The two spin branches are degenerate, and the baselines match the complete displaced-oscillator spectrum over the full scan.
		}
		\label{fig:Fig1}
	\end{figure}

	The pencil matrix components in the identity--identity corner,
	\begin{align}
		\hat{C}_{\mathrm{II}} =&~ c_{0}\hat{I}_{2}, \qquad
		\hat{D}_{\mathrm{II}} = d_{0}\hat{I}_{2},
	\end{align}
	provide a scalar standard matrix with null discriminant and vanishing nilpotent part,
	\begin{align}
		\hat{M}_{\mathrm{II}} =&~ \frac{c_{0}}{d_{0}}\hat{I}_{2}, \qquad
		\Delta_{\mathrm{M}} = 0, \qquad
		\hat{J} = 0.
	\end{align}
	The pencil factor leads to driven oscillators decoupled from the spin,
	\begin{align}
		\begin{aligned}
			\hat{A}_{\mathrm{II}} =&~ c_{0} + d_{0}\hat{a}, \\
			\hat{\mathcal{H}}_{\mathrm{II}-} =&~ d_{0}^{2}\hat{a}^{\dagger}\hat{a} + c_{0}d_{0}\left( \hat{a} + \hat{a}^{\dagger} \right) + c_{0}^{2}, \\
			\hat{\mathcal{H}}_{\mathrm{II}+} =&~ d_{0}^{2}\hat{a}^{\dagger}\hat{a} + c_{0}d_{0}\left( \hat{a} + \hat{a}^{\dagger} \right) + c_{0}^{2} + d_{0}^{2}.
		\end{aligned}
	\end{align}
	The zero-mode equation for the normal product gives a spin-degenerate coherent ground doublet,
	\begin{align}
		\lvert \psi_{\mathrm{II},0}^{(\pm)} \rangle =&~ \left\vert -\mu_{\mathrm{II}} \right\rangle \otimes \lvert m_{\mathrm{II},\pm} \rangle,
	\end{align}
	using a common displacement parameter and a basis for the spectator spin,
	\begin{align}
		\mu_{\mathrm{II}} =&~ \frac{c_{0}}{d_{0}},    \qquad
		\lvert m_{\mathrm{II},+} \rangle = \left( 1,0 \right)^{\mathrm{T}}, \qquad
		\lvert m_{\mathrm{II},-} \rangle = \left( 0,1 \right)^{\mathrm{T}}.
	\end{align}
	Both spin branches inherit the oscillator spacing in their normal-product baselines,
	\begin{align}
		E_{N}^{(\mathrm{II},\pm)} =&~ N\gamma_{\mathrm{II},\pm}, \qquad
		\gamma_{\mathrm{II},\pm} = d_{0}^{2}, \qquad
		N \in \mathbb{N}_{0}.
	\end{align}
	Although the two coherent gauges coalesce here, the baseline slope remains insensitive to the nonunique branch split, since $\hat{D}_{\mathrm{II}}^{\dagger} \hat{D}_{\mathrm{II}} = d_{0}^{2} \hat{I}_{2}$ gives
	$\operatorname{Tr}\!\left[\hat{P}\left(\hat{D}_{\mathrm{II}}^{\dagger}\hat{D}_{\mathrm{II}}\right)^{-1}\right]=d_{0}^{-2}$ for every rank-one projector $\hat{P}$.
	Figure~\ref{fig:Fig1} shows the exact finite-Fock spectrum against the coherent-gauge baselines across the identity--identity scan.
	
	These baselines match the complete driven-oscillator spectrum exactly, with a twofold spin degeneracy at every rung.
	For $N>0$, displacing the number states gives the exact bosonic excited states,
	\begin{align}
		\hat{D}_{\mathrm{b}}\left( -\mu_{\mathrm{II}} \right)\lvert N\rangle,
	\end{align}
	with Bargmann representation
	\begin{align}
		\psi_{N}(z) =&~ \frac{1}{\sqrt{N!}} e^{-\frac{1}{2}\mu_{\mathrm{II}}^{2}} e^{-\mu_{\mathrm{II}}z} \left( z + \mu_{\mathrm{II}} \right)^{N}.
	\end{align}
	This is exactly the degree-$N$ finite-polynomial dressing of the coherent gauge $e^{-\mu_{\mathrm{II}}z}$, with nonzero top coefficient $1/\sqrt{N!}$.

	\subsection{Exchange--identity pencil for the constrained anisotropic Rabi model}
	\label{sec:Sec52}
	
\begin{figure}[t]
		\centering
		\includegraphics[scale=1]{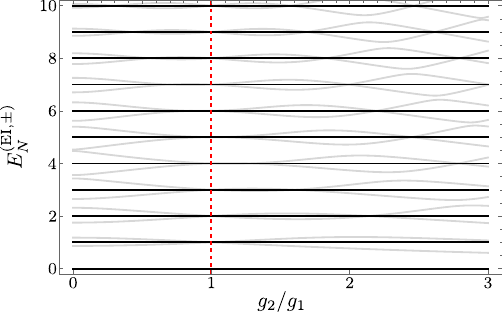}
		\caption{
			Exchange--identity spectral branch.
			Light gray curves show the numerical finite-Fock spectrum on $\Sigma_{+}$ for cutoff $N_{\mathrm{F}}=1000$, and solid black curves show the coherent-gauge baselines $\left. E_{N}^{(\mathrm{EI},\pm)}/\hbar \right\rvert_{\Sigma_{+}} = N\omega - (g_{1}^{2} + g_{2}^{2})/(2\omega)$.
			The red dashed line marks the symmetric point $g_{2}/g_{1}=1$, where the sheet constraint gives $\omega_{0}=0$ and the model reduces to the degenerate-qubit isotropic Rabi model.
		}
		\label{fig:Fig2}
	\end{figure}

	The pencil matrix components in the exchange--identity corner,
	\begin{align}
		\hat{C}_{\mathrm{EI}} =&~ \frac{g_{1}}{\omega}\hat{\sigma}_{-} + \frac{g_{2}}{\omega}\hat{\sigma}_{+},
		\qquad
		\hat{D}_{\mathrm{EI}} = \hat{I}_{2},
	\end{align}
	provide a real-split standard matrix for $g_{1}g_{2}>0$,
	\begin{align}
		\hat{M}_{\mathrm{EI}} =&~ \frac{1}{\omega}\left( g_{1}\hat{\sigma}_{-} + g_{2}\hat{\sigma}_{+} \right),
		\qquad
		\Delta_{\mathrm{M}}^{(\mathrm{EI})} = \frac{4g_{1}g_{2}}{\omega^{2}}.
	\end{align}
	The pencil factor leads to the Tomka--Pletyukhov--Gritsev supersymmetric factorization sheet~\cite{Tomka2015p13097},
	\begin{align}
		\begin{aligned}
			\hat{A}_{\mathrm{EI}} =&~ \hat{a} + \frac{1}{\omega}\left( g_{1}\hat{\sigma}_{-} + g_{2}\hat{\sigma}_{+} \right), \\
			\omega\hat{\mathcal{H}}_{\mathrm{EI}-} =&~
			\left. \frac{\hat{H}_{\mathrm{AR}}}{\hbar} \right\rvert_{\Sigma_{+}}
			+ \frac{g_{1}^{2} + g_{2}^{2}}{2\omega}, \\
			\omega\hat{\mathcal{H}}_{\mathrm{EI}+} =&~
			\left. \frac{\hat{H}_{\mathrm{AR}}}{\hbar} \right\rvert_{\Sigma_{-}}
			+ \frac{g_{1}^{2} + g_{2}^{2}}{2\omega}
			+ \omega,
		\end{aligned}
	\end{align}
	on the ordered-product sheets
	\begin{align}
		\Sigma_{\pm}: \qquad
		\omega\omega_{0} =&~ \pm\left( g_{1}^{2} - g_{2}^{2} \right).
	\end{align}
	The zero-mode equation for the normal product gives a coherent-spinor ground doublet,
	\begin{align}
		\lvert \psi_{\mathrm{EI},0}^{(\pm)} \rangle =&~
		\left\vert -\mu_{\mathrm{EI},\pm} \right\rangle
		\otimes
		\lvert m_{\mathrm{EI},\pm} \rangle,
	\end{align}
	with opposite displacement parameters and their spin eigenvectors,
	\begin{align}
		\mu_{\mathrm{EI},\pm} =&~ \pm\frac{\sqrt{g_{1}g_{2}}}{\omega},
		\qquad
		\lvert m_{\mathrm{EI},\pm} \rangle =
		\frac{1}{\sqrt{g_{1} + g_{2}}}
		\left( \sqrt{g_{2}}, \pm\sqrt{g_{1}} \right)^{\mathrm{T}}.
	\end{align}
	The normal-product zero energy maps to the anisotropic-Rabi energy on $\Sigma_{+}$,
	\begin{align}
		\frac{E_{\mathrm{EI},0}}{\hbar} =&~
		-\frac{g_{1}^{2} + g_{2}^{2}}{2\omega}.
	\end{align}
	The identity spin metric gives unit slopes for both normal-product baseline branches,
	\begin{align}
		E_{N}^{(\mathrm{EI},\pm)} =&~ N\gamma_{\mathrm{EI},\pm},
		\qquad
		\gamma_{\mathrm{EI},\pm} = 1,
		\qquad
		N \in \mathbb{N}_{0}.
	\end{align}
	Multiplying by the scale $\omega$ and subtracting the normal-product shift expresses the baselines in the anisotropic-Rabi energy convention,
	\begin{align}
		\left. \frac{E_{N}^{(\mathrm{EI},\pm)}}{\hbar} \right\rvert_{\Sigma_{+}} =&~
		N\omega - \frac{g_{1}^{2} + g_{2}^{2}}{2\omega}.
	\end{align}
	Figure~\ref{fig:Fig2} compares the finite-Fock spectrum on the exchange--identity sheet with the scaled coherent-gauge baselines.
	
	For this sheet, the first two endpoint constraints reduce, up to nonzero overall factors, to
	\begin{align}
		\begin{aligned}
			P_{1}^{(\mu_{\pm})} \propto&~ u^{2} - v^{2}, \\
			P_{2}^{(\mu_{\pm})} \propto&~ \left( u^{2} - v^{2} \right) \left( u^{2} + v^{2} - 1 \right),
		\end{aligned}
	\end{align}
	in terms of the dimensionless exchange couplings
	\begin{align}
		u =&~ \frac{g_{1}}{\omega}, \qquad
		v = \frac{g_{2}}{\omega}.
	\end{align}
	The sheet condition on $\Sigma_{+}$ relates their common factor to the qubit detuning,
	\begin{align}
		u^{2} - v^{2} =&~ \frac{\omega_{0}}{\omega},
	\end{align}
	so the symmetric point $g_{1}=g_{2}$ populates every baseline, while away from that point the first excited rung remains empty and the circle
	\begin{align}
		g_{1}^{2} + g_{2}^{2} =&~ \omega^{2}
	\end{align}
	populates the second rung.
	This ladder agrees with the exceptional baselines of the anisotropic quantum Rabi model~\cite{Xie2014p021046,Chen2021p043708}, and with the variational cat-state description~\cite{Zhang2016p063824} and cat-state generation scheme~\cite{Wang2018p053061}.
	
	At the symmetric point $g_{1}=g_{2}=g$, the sheet constraint gives $\omega_{0}=0$, and the Hamiltonian reduces to the degenerate-qubit isotropic Rabi model,
	\begin{align}
		\left. \frac{\hat{H}_{\mathrm{AR}}}{\hbar} \right\rvert_{g_{1}=g_{2}} =&~
		\omega \hat{a}^{\dagger} \hat{a} + g \left( \hat{a}^{\dagger} + \hat{a} \right)\left( \hat{\sigma}_{+} + \hat{\sigma}_{-} \right).
	\end{align}
	A fixed rotation to the $\hat{\sigma}_{x}$ basis separates this symmetric point into two driven oscillators with displacements $\mp g/\omega$.
	Every baseline is populated there, and the exact excited states are displaced number states in the rotated spin branches, in agreement with the degenerate-qubit solution~\cite{MaldonadoVillamizar2019p013811}.
	
	\subsection{Identity--exchange pencil for the constrained anisotropic Rabi--Stark model}
	\label{sec:Sec53}
	
	The pencil matrix components in the identity--exchange corner,
	\begin{align}
		\hat{C}_{\mathrm{IE}} =&~ \frac{\alpha_{0}}{\sqrt{\omega}}\hat{I}_{2}, \qquad
		\hat{D}_{\mathrm{IE}} = \frac{\alpha_{+}}{\sqrt{\omega}}\hat{\sigma}_{+} + \frac{\alpha_{-}}{\sqrt{\omega}}\hat{\sigma}_{-},
	\end{align}
	provide a real-split standard matrix for $\alpha_{+}\alpha_{-}>0$,
	\begin{align}
		\hat{M}_{\mathrm{IE}} =&~ \hat{D}_{\mathrm{IE}}^{-1}\hat{C}_{\mathrm{IE}}, \qquad
		\Delta_{\mathrm{M}}^{(\mathrm{IE})} = \frac{4\alpha_{0}^{2}\alpha_{+}\alpha_{-}}{\omega^{2}}.
	\end{align}
	The pencil factor leads to the constrained anisotropic Rabi--Stark factorization sheet,
	\begin{align}
		\begin{aligned}
			\hat{A}_{\mathrm{IE}} =&~ \frac{\alpha_{0}}{\sqrt{\omega}}\hat{I}_{2} + \frac{\hat{a}}{\sqrt{\omega}}\left( \alpha_{+}\hat{\sigma}_{+} + \alpha_{-}\hat{\sigma}_{-} \right), \\
			\omega\hat{\mathcal{H}}_{\mathrm{IE}-} =&~ \left. \frac{\hat{H}_{\mathrm{ARS}}}{\hbar} \right\rvert_{\Sigma_{\mathrm{K},-}} + \alpha_{0}^{2}, \\
			\omega\hat{\mathcal{H}}_{\mathrm{IE}+} =&~ \left. \frac{\hat{H}_{\mathrm{ARS}}}{\hbar} \right\rvert_{\Sigma_{\mathrm{K},+}} + \alpha_{0}^{2} + \omega.
		\end{aligned}
	\end{align}
	Matching the ordered products to the Rabi--Stark Hamiltonian constrains the couplings and Stark shift,
	\begin{align}
		\Sigma_{\mathrm{K},\pm}: \qquad
		\frac{g_{1}^{2}}{\omega + \omega_{0}} =&~ \frac{g_{2}^{2}}{\omega - \omega_{0}} = \alpha_{0}^{2}, \qquad
		\chi = \pm\omega_{0},
	\end{align}
	with the frequencies and exchange couplings expressed through the pencil parameters,
	\begin{align}
		\begin{aligned}
			\omega + \omega_{0} =&~ \alpha_{+}^{2}, \qquad
			\omega - \omega_{0} = \alpha_{-}^{2}, \\
			g_{1} =&~ \alpha_{0}\alpha_{+}, \qquad
			g_{2} = \alpha_{0}\alpha_{-}.
		\end{aligned}
	\end{align}
	This corner realizes a constrained anisotropic Rabi--Stark sheet~\cite{Eckle2017p294004,Xie2019p245304}.
	
\begin{figure}[t]
		\centering
		\includegraphics[scale=1]{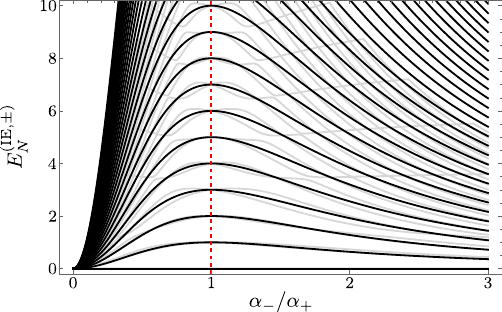}
		\caption{
			Identity--exchange spectral branch.
			Light gray curves show the numerical finite-Fock spectrum on $\Sigma_{\mathrm{K},-}$ for cutoff $N_{\mathrm{F}}=1000$, and solid black curves show the coherent-gauge baselines $\left. E_{N}^{(\mathrm{IE},\pm)}/\hbar \right\rvert_{\Sigma_{\mathrm{K},-}} = N(\omega^{2} - \omega_{0}^{2})/\omega - \alpha_{0}^{2}$.
			The red dashed line marks the symmetric point $\alpha_{-}/\alpha_{+}=1$, where the identity--exchange and exchange--identity sheets overlap at the degenerate-qubit isotropic Rabi model.
		}
		\label{fig:Fig3}
	\end{figure}

	The zero-mode equation for the normal product gives a coherent-spinor ground doublet,
	\begin{align}
		\lvert \psi_{\mathrm{IE},0}^{(\pm)} \rangle =&~ \left\vert -\mu_{\mathrm{IE},\pm} \right\rangle \otimes \lvert m_{\mathrm{IE},\pm} \rangle,
	\end{align}
	with displacement parameters and spin eigenvectors determined by the pencil coefficients,
	\begin{align}
		\mu_{\mathrm{IE},\pm} =&~ \pm \frac{\alpha_{0}}{\sqrt{\alpha_{+}\alpha_{-}}}, \qquad
		\lvert m_{\mathrm{IE},\pm} \rangle = \frac{1}{\sqrt{\alpha_{+} + \alpha_{-}}} \left( \sqrt{\alpha_{+}}, \pm\sqrt{\alpha_{-}} \right)^{\mathrm{T}}.
	\end{align}
	The normal-product zero energy maps to the Rabi--Stark energy on $\Sigma_{\mathrm{K},-}$,
	\begin{align}
		\frac{E_{\mathrm{IE},0}}{\hbar} =&~ -\alpha_{0}^{2}.
	\end{align}
	The inverse spin metric gives equal slopes for the two normal-product baseline branches,
	\begin{align}
		E_{N}^{(\mathrm{IE},\pm)} =&~ N\gamma_{\mathrm{IE},\pm}, \qquad
		\gamma_{\mathrm{IE},\pm} = \frac{\omega^{2} - \omega_{0}^{2}}{\omega^{2}}, \qquad
		N \in \mathbb{N}_{0}.
	\end{align}
	Multiplying by the scale $\omega$ and subtracting the normal-product shift expresses the baselines in the Rabi--Stark energy convention,
	\begin{align}
		\left. \frac{E_{N}^{(\mathrm{IE},\pm)}}{\hbar} \right\rvert_{\Sigma_{\mathrm{K},-}} =&~
		N\frac{\omega^{2} - \omega_{0}^{2}}{\omega} - \alpha_{0}^{2}.
	\end{align}
	Figure~\ref{fig:Fig3} compares the finite-Fock spectrum on the identity--exchange sheet with the scaled coherent-gauge baselines.
	
	The invertibility of $\hat{D}_{\mathrm{IE}}$ excludes the collapse boundaries $\omega=\pm\omega_{0}$~\cite{Xie2020p053803,Braak2024pC97}.
	For generic parameters on $\Sigma_{\mathrm{K},-}$, the coherent-gauge baselines provide the exceptional reference structure selected by the endpoint constraints.
	The peaked baseline shape in Fig.~\ref{fig:Fig3} follows from $\gamma_{\mathrm{IE},\pm}=(\omega^{2}-\omega_{0}^{2})/\omega^{2}$, which vanishes at those excluded boundaries.
	The equal branch slopes reproduce the equally spaced two-branch structure of the constrained Rabi--Stark spectrum.
	Our normal-product factorization has a two-dimensional zero-mode kernel, whereas our earlier factorization selects a unique ground state~\cite{Kafuri2024pC82}.
	This difference reflects the distinct factor operators and zero-mode conditions.
	At the symmetric point $\alpha_{+}=\alpha_{-}$, the sheet constraints give $\omega_{0}=0$, $\chi=0$, and $g_{1}=g_{2}$.
	The identity--exchange and exchange--identity sheets overlap at the degenerate-qubit isotropic Rabi model, with $g=\alpha_{0}\alpha_{+}$.
	The same rotation to the $\hat{\sigma}_{x}$ basis separates the Hamiltonian into two driven oscillators, so every baseline is populated there by displaced number states~\cite{MaldonadoVillamizar2019p013811}.
	
	\subsection{Exchange--exchange pencil in the longitudinal spin--boson limit}
	\label{sec:Sec54}
	
	The pencil matrix components in the exchange--exchange corner,
	\begin{align}
		\hat{C}_{\mathrm{EE}} =&~ c_{-}\hat{\sigma}_{-} + c_{+}\hat{\sigma}_{+}, \qquad
		\hat{D}_{\mathrm{EE}} = d_{-}\hat{\sigma}_{-} + d_{+}\hat{\sigma}_{+},
	\end{align}
	provide a diagonal standard matrix in the $\hat{\sigma}_{z}$ basis,
	\begin{align}
		\hat{M}_{\mathrm{EE}} =&~ \hat{D}_{\mathrm{EE}}^{-1}\hat{C}_{\mathrm{EE}} =
		\begin{pmatrix}
			\frac{c_{-}}{d_{-}} & 0 \\
			0 & \frac{c_{+}}{d_{+}}
		\end{pmatrix}, \qquad
		\Delta_{\mathrm{M}}^{(\mathrm{EE})} = \left( c_{-}d_{+} - c_{+}d_{-} \right)^{2}.
	\end{align}
	The diagonal entries determine the squared separation of the two standard-matrix eigenvalues,
	\begin{align}
		\left( \mu_{\mathrm{EE},+} - \mu_{\mathrm{EE},-} \right)^{2} =&~ \frac{\Delta_{\mathrm{M}}^{(\mathrm{EE})}}{\left( d_{+}d_{-} \right)^{2}} = \left( \frac{c_{-}}{d_{-}} - \frac{c_{+}}{d_{+}} \right)^{2}.
	\end{align}
	The exchange--exchange discriminant is a perfect square, so this corner realizes the real-split sector and its repeated-root boundary.
	The standard matrix becomes scalar at
	\begin{align}
		c_{-}d_{+} =&~ c_{+}d_{-},
	\end{align}
	while the two $\hat{\sigma}_{z}$-conditioned oscillator sectors remain physically well defined.
	
	The pencil factor leads to two driven oscillators conditioned by the spin-inversion charge,
	\begin{align}
		\begin{aligned}
			\hat{A}_{\mathrm{EE}} =&~ \left( c_{-} + d_{-}\hat{a} \right)\hat{\sigma}_{-} + \left( c_{+} + d_{+}\hat{a} \right)\hat{\sigma}_{+}, \\
			\hat{\mathcal{H}}_{\mathrm{EE}-} =&~ \frac{1}{2} \left[ \left( d_{-}^{2} + d_{+}^{2} \right)\hat{a}^{\dagger}\hat{a} + \left( c_{-}d_{-} + c_{+}d_{+} \right)\left( \hat{a} + \hat{a}^{\dagger} \right) + c_{-}^{2} + c_{+}^{2} \right]\hat{I}_{2} \\
			&~ + \frac{1}{2} \left[ \left( d_{-}^{2} - d_{+}^{2} \right)\hat{a}^{\dagger}\hat{a} + \left( c_{-}d_{-} - c_{+}d_{+} \right)\left( \hat{a} + \hat{a}^{\dagger} \right) + c_{-}^{2} - c_{+}^{2} \right]\hat{\sigma}_{z}, \\
			\hat{\mathcal{H}}_{\mathrm{EE}+} =&~ \frac{1}{2} \left[ \left( d_{-}^{2} + d_{+}^{2} \right)\hat{a}^{\dagger}\hat{a} + \left( c_{-}d_{-} + c_{+}d_{+} \right)\left( \hat{a} + \hat{a}^{\dagger} \right) + c_{-}^{2} + c_{+}^{2}
			+ d_{-}^{2} + d_{+}^{2} \right]\hat{I}_{2} \\
			&~ - \frac{1}{2} \left[ \left( d_{-}^{2} - d_{+}^{2} \right)\hat{a}^{\dagger}\hat{a} + \left( c_{-}d_{-} - c_{+}d_{+} \right)\left( \hat{a} + \hat{a}^{\dagger} \right) + c_{-}^{2} - c_{+}^{2} + d_{-}^{2} - d_{+}^{2} \right] \hat{\sigma}_{z}.
		\end{aligned}
	\end{align}
	The exchange couplings vanish identically, $g_{1}=g_{2}=0$, so this corner realizes a qubit-state-conditioned oscillator structure exactly, not as a dispersive approximation~\cite{Richer2016p134501,Potts2025p153603}.
	
\begin{figure}[t]
		\centering
		\includegraphics[scale=1]{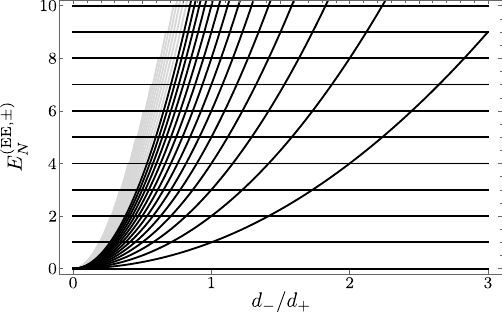}
		\caption{
			Exchange--exchange spectral branches.
			Light gray curves show the numerical finite-Fock spectrum of $\hat{\mathcal{H}}_{\mathrm{EE}-}$ for cutoff $N_{\mathrm{F}}=1000$, and solid black curves show the coherent-gauge baselines $E_{N}^{(\mathrm{EE},\pm)}=N d_{\mp}^{2}$.
			Each spin branch reduces to an independent scalar driven oscillator, so the baselines match the complete spin-conditioned spectrum over the full scan.
		}
		\label{fig:Fig4}
	\end{figure}

	The zero-mode equation for the normal product gives independently displaced coherent ground states,
	\begin{align}
		\lvert \psi_{\mathrm{EE},0}^{(\pm)} \rangle =&~
		\left\vert -\mu_{\mathrm{EE},\pm} \right\rangle
		\otimes
		\lvert m_{\mathrm{EE},\pm} \rangle,
	\end{align}
	with branch displacement parameters and the spin-inversion eigenvectors,
	\begin{align}
		\mu_{\mathrm{EE},\pm} =&~ \frac{c_{\mp}}{d_{\mp}}, \qquad
		\lvert m_{\mathrm{EE},+} \rangle = \left( 1,0 \right)^{\mathrm{T}}, \qquad
		\lvert m_{\mathrm{EE},-} \rangle = \left( 0,1 \right)^{\mathrm{T}}.
	\end{align}
	Each spin-conditioned oscillator supplies its own normal-product baseline spacing,
	\begin{align}
		E_{N}^{(\mathrm{EE},\pm)} =&~ N\gamma_{\mathrm{EE},\pm},
		\qquad
		\gamma_{\mathrm{EE},\pm} = d_{\mp}^{2},
		\qquad
		N \in \mathbb{N}_{0}.
	\end{align}
	Figure~\ref{fig:Fig4} shows the exact finite-Fock spectrum against the two coherent-gauge baseline branches across the exchange--exchange scan.
	The small-ratio edge approaches the regular-pencil boundary $d_{-}=0$, so the numerical spectrum starts slightly inside the invertible domain while the analytic baselines extend to the boundary.
	
	The two baseline branches coincide only for $d_{+}^{2} = d_{-}^{2}$.
	Since $\hat{M}_{\mathrm{EE}}$ is diagonal, each spin branch reduces to an independent scalar driven oscillator of the identity--identity type.
	These baselines match the complete spin-conditioned driven-oscillator spectrum exactly.
	Every rung is populated without endpoint constraints, and displacing the number states gives the exact excited states,
	\begin{align}
		\lvert \psi_{\mathrm{EE},N}^{(\pm)} \rangle =&~ \hat{D}_{\mathrm{b}}\left( -\mu_{\mathrm{EE},\pm} \right) \lvert N\rangle \otimes \lvert m_{\mathrm{EE},\pm} \rangle.
	\end{align}
	The branch with $\lvert m_{\mathrm{EE},+} \rangle$ has spacing $d_{-}^{2}$, while the branch with $\lvert m_{\mathrm{EE},-} \rangle$ has spacing $d_{+}^{2}$.
	Their splitting is a structural prediction of the regular spin-pencil class.
	
	\subsection{Interior pencil for a driven generalized Rabi--Stark model}
	\label{sec:Sec55}
	
	The sector-square corners isolate limiting factorizations.
	We propose a minimal interior example that keeps the exchange direction of the exchange--identity corner and populates the identity and exchange sectors of both matrices,
	\begin{align}
		\hat{X} =&~ \frac{g_{1}}{\omega}\hat{\sigma}_{-} + \frac{g_{2}}{\omega}\hat{\sigma}_{+}, \qquad
		\hat{C}_{\mathrm{CD}} = c_{0}\hat{I}_{2} + \hat{X}, \qquad
		\hat{D}_{\mathrm{CD}} = \hat{I}_{2} + \eta\hat{X}.
	\end{align}
	We choose parameters satisfying the interior and invertibility conditions
	\begin{align}
		c_{0} \eta \neq 0, \qquad
		g_{1}g_{2} > 0, \qquad    
		c_{0}^{2} - \frac{g_{1}g_{2}}{\omega^{2}} \neq 0, \qquad
		1 - \eta^{2}\frac{g_{1}g_{2}}{\omega^{2}} \neq 0,
	\end{align}
	so each invertible spin matrix has nonzero identity and exchange components.
	The affine factor defines generalized driven anisotropic Rabi--Stark Hamiltonians,
	\begin{align}
		\begin{aligned}
			\hat{A}_{\mathrm{CD}} =&~ c_{0} + \frac{g_{1}}{\omega} \hat{\sigma}_{-} + \frac{g_{2}}{\omega} \hat{\sigma}_{+} + \hat{a} + \eta \frac{g_{1}}{\omega}\hat{R}_{+} + \eta\frac{g_{2}}{\omega} \hat{Q}_{+}, \\
			\hat{\mathcal{H}}_{\mathrm{CD}-} =&~ \hat{A}_{\mathrm{CD}}^{\dagger} \hat{A}_{\mathrm{CD}}, \\
			\hat{\mathcal{H}}_{\mathrm{CD}+} =&~ \hat{A}_{\mathrm{CD}}\hat{A}_{\mathrm{CD}}^{\dagger}.
		\end{aligned}
	\end{align}
	These members are interior but remain aligned with the exchange--identity spin direction because $\hat{C}_{\mathrm{CD}}$ and $\hat{D}_{\mathrm{CD}}$ are polynomials in the same exchange matrix $\hat{X}$.
	Choosing a positive scale $\Omega_{\mathrm{CD}}$, the resulting driven generalized anisotropic Rabi--Stark Hamiltonians contain a direct bosonic drive, transverse spin bias, longitudinal spin-dependent bosonic drive, Stark deformation, and intensity-dependent transverse spin drive, with frequencies
	\begin{widetext}
		\begin{align}
			\begin{aligned}
				\omega_{\mathrm{CD}\pm} =&~ \Omega_{\mathrm{CD}}\left[ 1 + \frac{\eta^{2}}{4\omega^{2}} \left(g_{+}^{2}+g_{-}^{2}\right) \right], \quad &
				\omega_{0,\mathrm{CD}\pm} =&~ -\frac{\Omega_{\mathrm{CD}}}{\omega^{2}} \left( \pm 1+\frac{\eta^{2}}{2}\right)g_{+}g_{-}, \\
				g_{1,\mathrm{CD}\pm} =&~ \frac{\Omega_{\mathrm{CD}}}{2\omega} \left[ (1+\eta c_{0})g_{+} +(1-\eta c_{0})g_{-} \right], \quad &
				g_{2,\mathrm{CD}\pm} =&~ \frac{\Omega_{\mathrm{CD}}}{2\omega} \left[ (1+\eta c_{0})g_{+} -(1-\eta c_{0})g_{-} \right], \\
				\chi_{\mathrm{CD}\pm} =&~ \mp\frac{\eta^{2}\Omega_{\mathrm{CD}}}{2\omega^{2}}
				g_{+}g_{-}, \quad &
				\epsilon_{\mathrm{CD}\pm} =&~ \Omega_{\mathrm{CD}}\left[ c_{0} + \frac{\eta}{4\omega^{2}} \left(g_{+}^{2}+g_{-}^{2}\right) \right], \\
				\nu_{\mathrm{CD}\pm} =&~ \frac{\Omega_{\mathrm{CD}}}{\omega} \left[ c_{0} + \frac{1\pm1}{2}\eta \right]g_{+}, \quad &
				\lambda_{\mathrm{CD}\pm} =&~ \mp\frac{\eta\Omega_{\mathrm{CD}}}{2\omega^{2}}
				g_{+}g_{-}, \\
				\kappa_{\mathrm{CD}\pm} =&~ \frac{\eta\Omega_{\mathrm{CD}}}{\omega}g_{+},
				\quad &
				\varepsilon_{\mathrm{CD}\pm} =&~ \Omega_{\mathrm{CD}}\left[ c_{0}^{2} + \frac{1\pm1}{2} + \frac{ 1 + \frac{1 \pm 1}{2}\eta^{2} }{4\omega^{2}} \left(g_{+}^{2}+g_{-}^{2}\right) \right],
			\end{aligned}
		\end{align}
	\end{widetext}
	where we define the auxiliary couplings,
	\begin{align}
		g_{\pm} =&~ g_{1} \pm g_{2}.
	\end{align}
	
	The aligned pencil expresses the standard matrix as a rational function of the exchange matrix,
	\begin{align}
		\hat{M}_{\mathrm{CD}} =&~
		\left( \hat{I}_{2} + \eta\hat{X} \right)^{-1}
		\left( c_{0}\hat{I}_{2} + \hat{X} \right).
	\end{align}
	The pencil discriminant and standard-matrix roots,
	\begin{align}
		\Delta_{\mathrm{M}}^{(\mathrm{CD})} = 4s^{2}\left( 1 - \eta c_{0} \right)^{2}, \qquad
		\mu_{\mathrm{CD},\pm} = \frac{c_{0} \pm s}{1 \pm \eta s},
	\end{align}
	give the standard-matrix branch separation 
	\begin{align}
		\left( \mu_{\mathrm{CD},+} - \mu_{\mathrm{CD},-} \right)^{2} =&~
		\frac{4s^{2}\left( 1 - \eta c_{0} \right)^{2}}
		{\left( 1 - \eta^{2}s^{2} \right)^{2}},
	\end{align}
	where the dimensionless exchange scale combines the two couplings,
	\begin{align}
		s =&~ \frac{\sqrt{g_{1}g_{2}}}{\omega}.
	\end{align}
	The common exchange direction preserves the right spin eigenvectors of the exchange--identity corner,
	\begin{align}
		\lvert m_{\mathrm{CD},\pm} \rangle =&~
		\frac{1}{\sqrt{g_{1} + g_{2}}}
		\left( \sqrt{g_{2}}, \pm\sqrt{g_{1}} \right)^{\mathrm{T}},
	\end{align}
	and the zero-mode equation for the normal product gives the coherent-spinor ground doublet,
	\begin{align}
		\lvert \psi_{\mathrm{CD},0}^{(\pm)} \rangle =&~
		\left\vert -\mu_{\mathrm{CD},\pm} \right\rangle
		\otimes
		\lvert m_{\mathrm{CD},\pm} \rangle.
	\end{align}
	The identity component $c_{0}$ and the exchange component in $\hat{D}_{\mathrm{CD}}$ shift the coherent displacements from the exchange--identity values, while the aligned construction leaves the spin dressing unchanged.
	
\begin{figure}[t]
		\centering
		\includegraphics[scale=1]{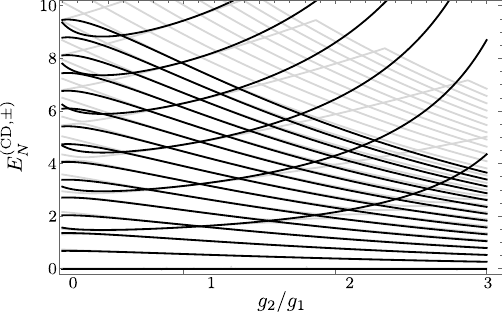}
		\caption{
			Driven interior spectral branches.
			Light gray curves show the numerical finite-Fock spectrum of $\hat{\mathcal{H}}_{\mathrm{CD}-}$ for cutoff $N_{\mathrm{F}}=1000$, and solid black curves show the coherent-gauge baselines $E_{N}^{(\mathrm{CD},\pm)}=N\gamma_{\mathrm{CD},\pm}$.
			Unlike the identity--identity and exchange--exchange corners, the interior baselines are not complete oscillator spectra; they provide the exceptional reference energies selected by the endpoint constraints.
		}
		\label{fig:Fig5}
	\end{figure}

	The inverse spin metric determines the slopes of the normal-product baselines,
	\begin{align}
		E_{N}^{(\mathrm{CD},\pm)} =&~ N\gamma_{\mathrm{CD},\pm}, \qquad
		\frac{1}{\gamma_{\mathrm{CD},\pm}} = \frac{\langle n_{\mathrm{CD},\pm}\rvert\left( \hat{D}_{\mathrm{CD}}^{\dagger}\hat{D}_{\mathrm{CD}} \right)^{-1}\lvert m_{\mathrm{CD},\pm} \rangle}{\langle n_{\mathrm{CD},\pm}\vert m_{\mathrm{CD},\pm} \rangle}, \qquad
		N \in \mathbb{N}_{0},
	\end{align}
	where $\langle n_{\mathrm{CD},\pm}\rvert$ are the left spin eigenvectors of $\hat{M}_{\mathrm{CD}}$.
	For $\eta\neq0$, the interior deformation generically splits the two branch slopes away from their exchange--identity value, even though the spin eigenvectors remain the mixed exchange--identity spinors.
	For the parameter scan in Fig.~\ref{fig:Fig5}, the branch slopes remain finite over the plotted interval.
	Figure~\ref{fig:Fig5} compares the finite-Fock spectrum of our aligned interior example with the coherent-gauge baselines.
	
	\section{Conclusion}
	\label{sec:Sec6}
	
	We constructed a regular spin-pencil factorization for spin-boson Hamiltonians generated by an affine first-order factor $\hat{A}=\hat{C}+\hat{a}\hat{D}$ with real invertible spin matrices $\hat{C}$ and $\hat{D}$.
	The normal product $\hat{A}^{\dagger}\hat{A}$ has an exact zero-energy ground doublet, while $\hat{A}\hat{A}^{\dagger}$ has no zero modes, giving a factor index of two.
	The standard matrix $\hat{M}=\hat{D}^{-1}\hat{C}$ determines the coherent displacements at distinct roots and the scalar or Jordan structure at repeated roots, where the two-dimensional zero-mode kernel persists.
	For distinct roots, the coherent gauges give finite baseline energies when their reciprocal slopes are nonzero, and the endpoint constraints determine which baselines support finite-polynomial eigenstates.
	
	The identity and exchange subspaces of $\hat{C}$ and $\hat{D}$ organize four limiting cases.
	The identity--identity corner reduces to a spin-degenerate driven oscillator, while the exchange--exchange corner gives two spin-conditioned driven oscillators with generically unequal spacings.
	Every baseline in these two corners is an exact spectral branch with displaced-number eigenstates.
	The exchange--identity corner recovers the Tomka--Pletyukhov--Gritsev anisotropic-Rabi sheet, while the identity--exchange corner gives our constrained anisotropic Rabi--Stark sheet.
	At their symmetric points, the mixed sheets meet at the degenerate-qubit isotropic Rabi model, where a fixed $\hat{\sigma}_{x}$ rotation separates the Hamiltonian into two driven oscillators and every baseline is exact.
	
	On the exchange--identity sheet, the first excited rung remains empty away from the symmetric point, while the second becomes exceptional on the circle $g_{1}^{2}+g_{2}^{2}=\omega^{2}$.
	Our aligned interior example retains the spin eigenvectors of the exchange--identity corner while changing the coherent displacements and branch slopes, and combines direct and longitudinal drives, Stark deformation, and intensity-dependent transverse spin coupling.
	
	Beyond this aligned example, the noncommuting interior remains to be explored systematically.
	Future work could examine complex pencil coefficients, non-Hermitian reservoir-induced terms, and higher-dimensional spin representations for multilevel and multimode generalizations.

	\section*{Funding}
	The authors received no external funding for this work.
	
	\begin{acknowledgments}
		B.~M.~R.~L. thanks Jacinta Alderete Galan for providing daycare support throughout this work.
		He also acknowledges support and hospitality as an affiliate visiting colleague at the Department of Physics and Astronomy, University of New Mexico.
	\end{acknowledgments}
	
	\section*{Disclosures}
	The authors declare no conflicts of interest.
	
	\section*{Data Availability Statement}
	The data that support the findings of this study are available from the corresponding author upon reasonable request.

	
	%
	
\end{document}